\documentclass[preprint,a4paper,prd,preprintnumbers,amssymb,superscriptaddress,nofootinbib]{revtex4}
\pdfoutput=1
\usepackage[english]{babel}
\usepackage[text={17.0cm,23.0cm}]{geometry}
\usepackage{graphicx,epsfig,graphics,verbatim}
\usepackage{mathrsfs,amsmath,amssymb}
\usepackage{slashed}
\usepackage{psfrag}
\usepackage{color}
\usepackage[normalem]{ulem}
\usepackage[utf8]{inputenc}
\usepackage{hyperref}
\usepackage{subfigure}

\def\lsim{\mathrel{\rlap{\lower4pt\hbox{\hskip1pt$\sim$}}
    \raise1pt\hbox{$<$}}}                
\def\gsim{\mathrel{\rlap{\lower4pt\hbox{\hskip1pt$\sim$}}
    \raise1pt\hbox{$>$}}}                

\begin{document} 
\preprint{ULB-TH/15-18}
\preprint{TUM-HEP 1026/15}

\title{Signatures from Scalar Dark Matter \\with a Vector-like Quark  Mediator}

\author{Federica Giacchino}
\email{federica.giacchino@ulb.ac.be}
\affiliation{Service de Physique Th\'eorique, CP225, Universit\'e Libre de Bruxelles, Bld du Triomphe, 1050 Brussels, Belgium}
\author{Alejandro Ibarra}
\email{ibarra@tum.de}
\affiliation{Physik-Department T30d, Technische Universit\"at M\"unchen, James-Franck-Stra\ss{}e, D-85748 Garching, Germany}
\author{Laura Lopez Honorez}
\email{llopezho@vub.ac.be}
\affiliation{Vrije Universiteit Brussel and The International Solvay Institutes, Pleinlaan 2, B-1050 Brussels, Belgium}
\author{Michel H.G.~Tytgat}
\email{mtytgat@ulb.ac.be}
\affiliation{Service de Physique Th\'eorique, CP225, Universit\'e Libre de Bruxelles, Bld du Triomphe, 1050 Brussels, Belgium}
\author{Sebastian Wild}
\email{sebastian.wild@ph.tum.de}
\affiliation{Physik-Department T30d, Technische Universit\"at M\"unchen, James-Franck-Stra\ss{}e, D-85748 Garching, Germany}

\date{\today}

\begin{abstract}
We present a comprehensive study of a model where the dark matter is composed of a singlet real scalar that couples to the Standard Model predominantly via a Yukawa interaction with a light quark and a colored vector-like fermion.
A distinctive feature of this scenario is that thermal freeze-out in the early universe may be driven by annihilation both into gluon pairs at one-loop ($gg$) and by virtual internal Bremsstrahlung of  a gluon ($q\bar q g$). 
Such a dark matter candidate may also be tested through direct and
indirect detection and at the LHC; viable candidates
have either a mass nearly degenerate with that of the fermionic mediator or a mass above about 2 TeV.
\end{abstract}

\maketitle
\section{Introduction}

In recent years,  dark matter in the form of Weakly Interacting Massive Particles (WIMPs)  has become the leading particle physics candidate. The most salient feature of WIMP dark matter is the prediction of a relic abundance via thermal freeze-out that is in rough agreement with the measured value,  when the WIMP annihilation cross section into Standard Model (SM) particles is of the order of 1 pb, thus suggesting parameters in the dark sector (WIMP mass and couplings) comparable to those in the electroweak sector. Furthermore, many particle physics models addressing the electroweak hierarchy problem contain, among many new states, a dark matter candidate with the characteristics of a WIMP. While the lack of experimental evidence for these extra states is disfavoring some of these schemes, the freeze-out mechanism still stands as one of the most natural and elegant mechanisms to explain the origin of the dark matter. 

An exciting feature of the WIMP dark matter paradigm, and arguably one of the reasons for its popularity, is the possibility of detecting experimental signals of processes induced by the weak coupling of the dark matter with the Standard Model particles. Numerous experiments are currently searching for dark matter particles, either through direct and indirect detection or at colliders, and are already providing fairly stringent limits on the rates of the  processes relevant for each search strategy. Unfortunately, the limits on the fundamental parameters of the model which can be derived from the null searches, as well as the complementarity among the various search strategies, are  highly model dependent. On the other hand, the main features of the dark matter phenomenology of a given model can be, in many instances, captured by considering only a subset of the new fields and new parameters, namely by considering ``simplified models''. 

The simplest among all the simplified WIMP models is to extend the Standard Model with a singlet real scalar, $S$, and with a discrete $Z_2$ symmetry, unbroken in the electroweak vacuum, under which the singlet scalar is odd while all the Standard Model fields are even~\cite{Silveira:1985rk,McDonald:1993ex,Burgess:2000yq,Patt:2006fw}, and which ensures the dark matter stability.\footnote{Other parity assignments are possible. For instance stability of scalar DM may be explained through embedding in $SO(10)$ using matter parity $P = (-)^{3(B-L)}$, which is a remnant of a gauge symmetry \cite{Kadastik:2009dj}. In this framework, SM fermions  are as usual in a $16_{\rm SM}$ (odd under $P$) while the SM Higgs is in $10_H$ (even). If $P$ is unbroken, a possibility is to embed scalar DM in a $16_{\rm DM}$, which contains a singlet scalar and a scalar doublet (aka inert doublet). Either one may play the role of DM \cite{Kadastik:2009dj}. Notice that  Yukawas of the form of Eq.~(\ref{eq:Lint}) may be obtained by coupling the $16_{\rm DM}$ to $16_{\rm SM}$ through fermions in the $10 \rightarrow 5 \oplus \bar 5$, which contains  a $SU(2)_L$ singlet vector-like quark, akin to the $d_R$,  as well as a leptonic $SU(2)_L$ doublet. Of course more work would be necessary to make this a phenomenologically viable framework. For recent works along this direction, see \cite{Frigerio:2009wf,Arbelaez:2015ila,Rodejohann:2015lca,Nagata:2015dma,Mambrini:2015vna,Heeck:2015qra}. }
The singlet scalar interacts with the Standard Model particles only through a renormalizable quartic coupling to the Brout-Englert-Higgs doublet (Higgs for short)
\begin{equation}
 {\cal L} \supset -\frac12 m_S^2 S^2 - \frac12 \lambda S^2 H^\dagger H,
\label{eq:LHp}
\end{equation}
thus the model only contains one new field and two new parameters, $m_S$ and $\lambda$; the phenomenology of this very simple scenario has been studied in many works (see for instance \cite{Goudelis:2009zz,Yaguna:2008hd,Gonderinger:2009jp,Profumo:2010kp,Djouadi:2011aa,Cline:2013gha,Duerr:2015aka}). 

In this paper we will investigate the impact on the dark matter phenomenology of extending this model by one vector-like fermionic field, $\psi$, also odd under the discrete $Z_2$ symmetry, and which couples to the singlet scalar $S$ and a right-handed SM fermion $f_R$ via a Yukawa interaction (the discussion for a left-handed SM fermion is completely analogous). The Lagrangian of the $Z_2$-odd fermionic sector reads:
\begin{equation}
{\cal L} \supset - y\, S\, \bar\psi f_R+ \bar \psi( i \slashed D -m_\psi) \psi + {\rm h.c.} 
\label{eq:Lint}
\end{equation}
The gauge invariance of the Yukawa term implies that the fermion must be a $SU(2)_L$ singlet and have the same hypercharge as the SM fermion it couples to. Besides, the vector-like nature of the fermion ensures the cancellation of the gauge anomalies.\footnote{Taking $\psi$ chiral instead requires introducing more fields, but is otherwise an obvious extension (with a possible caveat regarding the mass of the $\psi$, which may have to come from couplings to the Higgs and thus may not be arbitrarily large).} 
From a more fundamental perspective, notice that this model is also a simplified version of a scenario with large extra dimensions (see \cite{Bertone:2009cb} and references therein).

As we will argue, when the strength of the ``vector-like portal'' interaction is larger than the strength of the ``Higgs-portal'' interaction,  the phenomenology  of the model presents quite distinctive characteristics which make this model rather unique. The phenomenology of the  vector-like portal interaction has been partially discussed in, {\it e.g.}, \cite{Boehm:2003hm,Vasquez:2009kq,Bertone:2009cb,Perez:2013nra,Toma:2013bka,Giacchino:2013bta,Chang:2014tea,Ibarra:2014qma,Giacchino:2014moa}. The case in which $f_R$ is a light lepton was first discussed in \cite{Toma:2013bka,Giacchino:2013bta}. There, it was shown that the annihilation of $S$ through the portal of Eq.~(\ref{eq:Lint}) into two SM fermions is dominated,  in the limit $m_f \ll m_S$, by the d-wave, thus $\sigma v \propto v^4$, and that the  virtual internal Bremsstrahlung (VIB) in the annihilation may lead to rather intense gamma ray spectral features (see also \cite{Bergstrom:1989jr,Flores:1989ru,Bringmann:2007nk,Garny:2011ii,Bringmann:2012vr} for relevant literature on VIB in an analogous framework with Majorana dark matter). Further phenomenological aspects of this ``leptophilic'' scenario were studied in \cite{Ibarra:2014qma} and \cite{Giacchino:2014moa}. In particular \cite{Ibarra:2014qma}  analyzed constraints both from indirect detection through gamma rays and from colliders. 

In the present work, we will study the possibility that the dark matter couples to a light quark, assuming for simplicity minimal flavour violation.
An interesting feature of this  scenario is that, due to the d-wave suppression of the annihilation rate into a quark-antiquark pair,  the relic abundance is driven either by the annihilation of $S$ into two gluons~\footnote{The possibility that the annihilation into gluon pairs could drive the dark matter freeze-out was previously considered in \cite{Chu:2012qy},  using a model-independent effective operator framework. The present model provides an explicit UV realization of this scenario.} ($gg$) and into VIB of a gluon ($q\bar q g$), or by the co-annihilation of the colored $\psi$ particle. The latter process can be significantly affected by the exchange of gluons between the non-relativistic fermion mediators, a phenomenon that may lead to Sommerfeld enhancement of the annihilation cross-section. The correct description of the phenomenology of the model then requires the inclusion of higher-order effects, as well as non-perturbative effects. 

The fact that the fermionic mediator $\psi$ is colored also opens interesting possibilities for direct detection and collider searches \footnote{As the couplings we consider violate flavour symmetry, it is legitimate to ask whether our scenario may be made consistent with constraints on flavour changing neutral processes. This issue has been addressed in \cite{Garny:2014waa} where two mechanisms to suppress such processes (degeneracy and alignment) are discussed. As these mechanisms directly apply to our model, we do not repeat the argument but refer to \cite{Garny:2014waa} for more details.}. Specifically,  we consider the constraints set by the LUX~\cite{Akerib:2013tjd} experiment and the prospects for detection by XENON1T \cite{Aprile:2012zx}. To this end, we take into account the effective coupling at one-loop of the $S$ to gluons, recently calculated  in \cite{Hisano:2015bma}. Furthermore, the model can be tested at colliders through the production of the mediator particles $\psi$, which are both colored and electrically charged, and which subsequently decay producing a signature of two or more jets plus  missing transverse energy $E_T$. Altogether, direct detection and production of the fermionic mediator
at colliders put the strongest constraints on our scenario. Finally we
also consider indirect searches, in particular constraints on continuum gamma-ray signals from dwarf spheroidal
galaxies (dSphs) and gamma-ray spectral features, based on data from Fermi-LAT and H.E.S.S., and
on  anti-protons in cosmic rays, using the PAMELA
data. While overall less constraining,
these data sets will allow us to close a narrow region of the
parameter space that otherwise would be left open by current direct
and collider searches. Similar analyses applied to other dark matter
scenarios have been pursued in
Refs.~\cite{Garny:2013ama,Ibarra:2014vya,Kopp:2014tsa,Papucci:2014iwa,Ibarra:2015nca,Garny:2015wea}.

\bigskip
This article is organized as follows. We first determine in Sec~\ref{sec:viable-param-space} the dark matter relic abundance, taking into account higher-order processes involving gluons, as well as the Sommerfeld enhancement for co-annihilation of the fermionic mediator. Then, we study  in  Secs.~\ref{sec:direct-detect-constr}, ~\ref{sec:collider-constr} and ~\ref{sec:antipr-constr} the constraints from direct searches, collider searches and indirect searches, respectively. We finally draw our conclusions in Sec.~\ref{sec:concl}.

\section{Dark matter relic density}
\label{sec:viable-param-space}

As will be shown below, the relic dark matter abundance is determined either by the one-loop annihilation into gluon pairs ($gg$) or the VIB in the annihilation ($q\bar q g$), or by the co-annihilation with the colored partner. We first discuss the perturbative aspects of the annihilation, and then  the Sommerfeld effects in the co-annihilation processes associated to the exchange of gluons between particles in initial states.

\subsection{Perturbative analysis}

We will focus on the scenario where the vector-like portal interaction dominates over the Higgs portal interaction, namely we assume that the quartic coupling is small enough to play no role. In this case, the lowest order dark matter annihilation channel is $S S \rightarrow q \bar{q}$, with the annihilation cross-section given by
\begin{align}
\sigma v \simeq {3 y^4 \over 4 \pi  m_S^2(1+ r^2)^2}\left(\frac{ m_q^2}{m_S^2} - {2\over 3} {m_q^2 v^2 (1+ 2 r^2)\over m_S^2 (1+ r^2)^2}+ \frac{v^4}{15 \left( 1+r^2 \right)^2}\right) \,,
\label{eq:sigmav_SSqq}
\end{align}
where we have kept only the leading terms in an expansion in $m_q \ll m_S$ and relative velocity $v$. Here, $r \equiv m_\psi/m_S > 1$ denotes the mass ratio between the vector-like mediator and the dark matter particle. It follows that in the limit $m_q \rightarrow 0$, which is relevant for the case of dark matter coupling to a light quark, the annihilation of dark matter into a pair of quarks is $d$-wave suppressed \cite{Toma:2013bka,Giacchino:2013bta}. While the absence of the $s$-wave component in the limit $m_q \rightarrow 0$ is analogous to the well-known helicity suppression in the case of annihilation of a Majorana fermion pair into light fermions, the vanishing of the $p$-wave component is unique to the annihilation of two real scalar particles; see~\cite{Giacchino:2013bta} for a more detailed discussion.

\begin{figure}
\begin{center}
\includegraphics[scale=0.73]{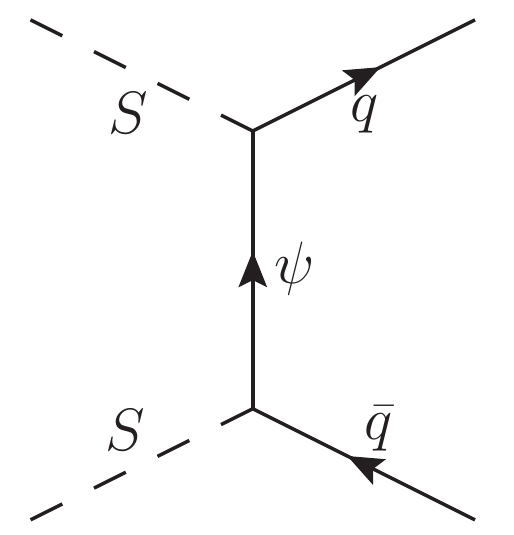}
\hspace{1.2cm}
\includegraphics[scale=0.73]{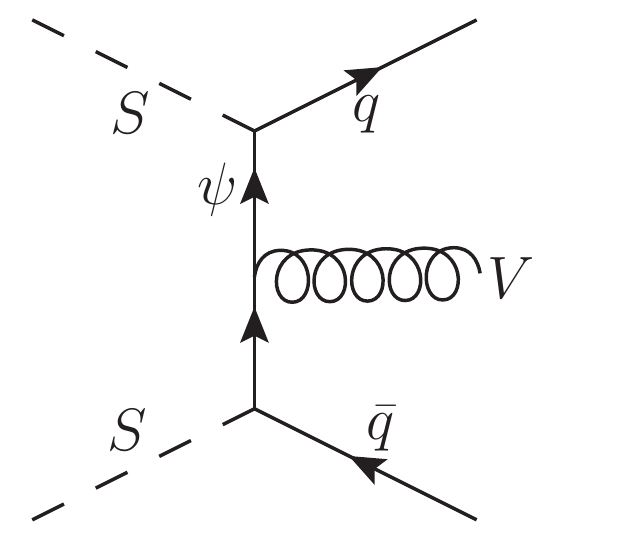}
\hspace{0.6cm}
\includegraphics[scale=0.73]{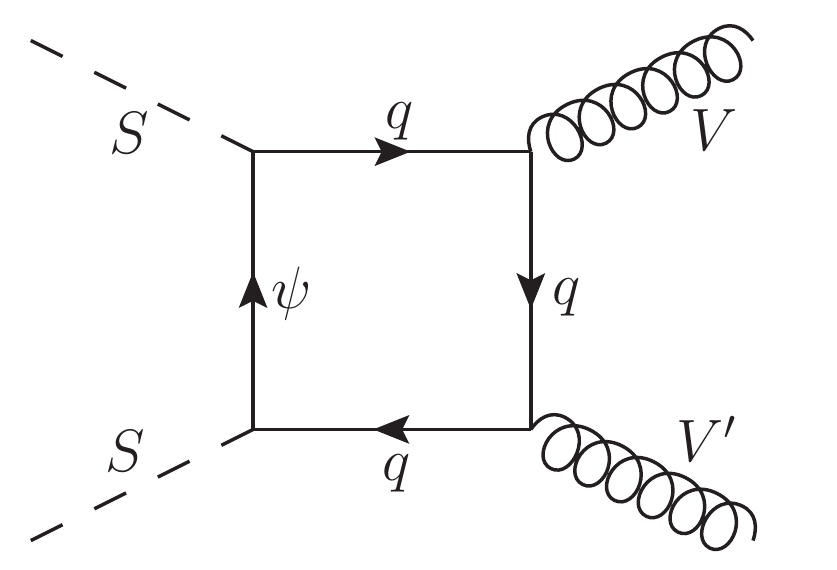}
\end{center}
\caption{\small Annihilation channels of the dark matter particle $S$ during freeze-out. In the limit $m_q \rightarrow 0$, the annihilation channel $SS \rightarrow q \bar q$, shown in the left panel, is $d$-wave suppressed (see text for details), and the higher order processes shown in the central and right panel dominate the total annihilation cross section. For each process, we only show one representative diagram.}
\label{fig:relic_density_diagrams}
\end{figure}

Due to the strong suppression of the tree-level two-to-two cross-section, it is crucial to take into account higher-order effects in the calculation of the relic density. There are two relevant classes of higher-order processes which contribute to the total annihilation cross section: $SS \rightarrow q \bar q V$, {\it i.e.} the internal Bremsstrahlung of a gauge boson $V$, with  $V = \gamma, Z, g$, as well as $SS \rightarrow V V'$, the one-loop annihilation of dark matter into a pair of gauge bosons $V, \,V'$. Representative diagrams for these higher-order annihilation channels are shown in Fig.~\ref{fig:relic_density_diagrams}, together with the tree-level annihilation process $SS\rightarrow q\bar{q}$. In contrast to Eq.~(\ref{eq:sigmav_SSqq}), the higher-order processes feature a non-vanishing $s$-wave component in the limit $m_q \rightarrow 0$, at the price of being suppressed by an additional gauge coupling and reduced phase space in the case of internal Bremsstrahlung, or by a loop suppression factor for annihilation into gauge bosons. Analytical expressions of the corresponding cross sections can be found in~\cite{Giacchino:2013bta,Toma:2013bka,Ibarra:2014qma,Giacchino:2014moa}.

To calculate the relic density, we employ the {\tt micrOMEGAS.4} package~\cite{Belanger:2014vza}, which we modified in order to include the additional annihilation channels.\footnote{For the higher-order processes, we only include the $s$-wave contribution to the annihilation cross section in the calculation of the relic density.} We find that in the whole parameter space either the internal Bremsstrahlung processes or the one-loop annihilations give a larger contribution to the total annihilation cross section at freeze-out than the annihilation into $q\bar q$ given by Eq.~(\ref{eq:sigmav_SSqq}). Due to the different scaling with the mass ratio $r$ of the two higher-order processes, namely $\sigma v \propto 1/r^8$ for internal Bremsstrahlung and $\sigma v \propto 1/r^4$ for one-loop annihilations (see~\cite{Giacchino:2013bta,Toma:2013bka,Ibarra:2014qma,Giacchino:2014moa}), the former dominates the total annihilation cross section for small mass ratios (concretely, $r \lesssim 2.5$), while the latter is more important for large mass ratios.

Furthermore, if the fermionic mediator is close in mass to the dark matter particle, co-annihilation processes, {\it i.e.}~annihilations of $S$ and $\psi$, $\psi$ and $\psi$, or $\psi$ and $\bar \psi$~\cite{Griest:1990kh}, can play a significant role in the calculation of the relic density. As is well known, the co-annihilation channels have rates exponentially suppressed when $\Delta m \equiv m_\psi - m_S \gg T_\text{f.o.}$, yet, for small mass splittings, they can dominate over the self-annihilation channels, which have in this model suppressed rates. We employ~{\tt micrOMEGAS.4} in order to include all relevant co-annihilation processes in our numerical calculation, and we find that co-annihilations give a contribution to $\Omega_\text{DM}$  of at least 5\% for mass splittings $r \lesssim 1.5,\,1.35,\,1.25$ at $m_S = 10\,\text{GeV},\,100\,\text{GeV},\,1\,\text{TeV}$, respectively. It is important to note that some of the co-annihilation processes as e.g. $\psi \bar{\psi} \rightarrow g g$ are purely given by gauge interactions; in particular, they are independent of the Yukawa coupling $y$. Consequently, if $m_S$ and $r$ are small enough, these annihilation channels can suppress the relic density below the observed value, irrespectively of the value of $y$.\footnote{Co-annihilation still requires that $S \leftrightarrow \psi$  transitions on SM particles of the thermal bath are in equilibrium at the time of freeze-out. We have checked that this is the case provided $y \gtrsim 10^{-4}$.}

For the parts of the parameter space where co-annihilations are not relevant, the relic density is set by the higher-order processes  introduced above, and consequently rather large Yukawa couplings are typically necessary for matching the observed relic density. As it will be discussed in the rest of this work, this implies potentially observable signals in direct and indirect detection, as well as in the production of the mediator at colliders.

\subsection{Sommerfeld corrections to the co-annihilation processes}
\label{subsec:sommerfeld}

The annihilation of two colored particles (for our purposes, with initial states $\psi \psi$, {$\bar\psi\bar\psi$ } or $\psi \bar \psi$) can be significantly affected by the non-perturbative Sommerfeld effect, induced by the multiple exchange of gluons in the initial state. For the quantitative treatment of the Sommerfeld effect, we closely follow the method presented in~\cite{deSimone:2014pda,Garny:2015wea}, which we briefly recapitulate here.

Expanding the perturbative annihilation cross section as
\begin{align}
  \sigma_\text{pert.}= \frac{a}{v}+b v+ {\cal O} (v^2)\,,
\end{align}
the Sommerfeld corrected cross section can be written as
\begin{align}
  \sigma_{\rm Somm.}= S_{0} \frac{a}{v}+ S_{1} b v+ {\cal O} (v^2)\,.
\end{align}
Here, $S_l$ is the Sommerfeld factor associated with the partial wave $l$, which takes the form~\cite{Cassel:2009wt,Iengo:2009ni}
\begin{align}
\label{eq:def_Sl}
S_{0} = \frac{-2 \pi \alpha/v}{1-\exp(2 \pi \alpha/v)} \quad , \quad S_{l>0} = S_0 \times \prod_{k=1}^l \left( 1+ \frac{\alpha^2}{v^2 k^2} \right) \,.
\end{align}
In these expressions, $v$ is the relative velocity between the two annihilating particles, and the coupling $\alpha$ parametrizes the strength of the effective QCD potential between the annihilating particles. For two annihilating particles, with representations $R$ and $R'$ under $SU(3)_c$, the non-abelian matrix potential between the particles can be diagonalized by decomposing the direct product $R\otimes R' =\sum_Q Q$ as a sum of irreducible representations $Q$. This gives rise to the effective potential (see e.g.~\cite{deSimone:2014pda})
\begin{equation}
  V(r) =\frac{\alpha}{r}=\frac{\alpha_s}{r}\frac12 \left(C_Q-C_R-C_{R'}\right) \,,
\label{eq:V}
\end{equation}
where we evaluate $\alpha_s$ at an energy scale equal to the momenta of the annihilating particles, $p=m v/2$ and the $C_i$ are the quadratic Casimir operators associated with the representation $i$. For the model under consideration, we will be interested in processes involving the annihilation of colored fermion triplets giving rise to $\textbf{3} \otimes \bar{\textbf{3}}= \textbf{1}\oplus \textbf{8}$ and $\textbf{3} \otimes \textbf{3}= \textbf{6}\oplus \bar{\textbf{3}}$. The associated Casimir operators are given by $ C_\textbf{1}=0,C_\textbf{3}=C_{\bar{\textbf{3}}} =4/3,C_\textbf{6}=10/3$ and $C_\textbf{8}=3$. The effective QCD potentials then read
\begin{equation}
  V_{\textbf{3}\otimes \bar{\textbf{3}}}=\frac{\alpha_s(\mu=p)}{r}
  \left\{
  \begin{array}{cc}
    -\frac43 & (\textbf{1})\\
    \frac16 & (\textbf{8})\\
  \end{array}
  \right.
 \qquad {\rm and}\qquad
 V_{\textbf{3}\otimes  \textbf{3}}=\frac{\alpha_s(\mu=p)}{r}
 \left\{
 \begin{array}{cc}
   -\frac23 & (\bar{\textbf{3}})\\
   \frac13 & (\textbf{6})\\
 \end{array}
 \right. \,,
\label{eq:VQ}
  \end{equation}
  and are hence attractive for the singlet and the anti-triplet two particles states while they are repulsive for the others. In the following, we will refer to $S_l^{(\textbf{1})}$, $S_l^{(\textbf{8})}$, $S_l^{(\bar{\textbf{3}})}$, $S_l^{(\textbf{6})}$ for the Sommerfeld factors defined in Eq.~(\ref{eq:def_Sl}), evaluated for a coupling $\alpha= \alpha_s \times \{ -4/3,1/6,-2/3,1/3\}$, respectively.

As the Sommerfeld enhancement depends on the initial color state, one has to determine the relative probabilities of annihilation in a given color state, separately for every annihilation channel. As discussed in~\cite{deSimone:2014pda}, using tensor decomposition,
one can show that $\psi \bar{\psi} \rightarrow gg$ occurs with probability 2/7 (5/7) through a singlet (octet) state, giving rise to the total Sommerfeld factor\footnote{These relative probabilities have been calculated in the limit $v\rightarrow 0$, {\it i.e.} for $S_{l>0}$ they are in principle (slightly) different. We have checked that the error associated to the simplification of applying the same probabilities also for $l>0$ is negligible. The same argument also holds for the other annihilation channels.}
\begin{equation}
  S^{(gg)}_l= \frac27 S_l^{(\textbf{1})}+\frac57 S_l^{(\textbf{8})}\,.
\label{eq:Sgg}
\end{equation}
For the annihilation channel $\psi \psi \rightarrow qq$ due to the
Yukawa interaction of Eq.~(\ref{eq:Lint}), a similar calculation
yields
\begin{equation}
S^{(qq)}_l = \frac13 S_l^{({\bf \bar 3})} + \frac23 S_l^{(\textbf{6})}  \,.
\label{eq:Sqq}
\end{equation}
On the other hand, the annihilations of $\psi \bar{\psi}$ into $\gamma g, Z g$ ($\gamma \gamma, \gamma Z, ZZ, WW, Zh$) correspond to a pure octet (singlet) initial state, due to color conservation. The associated Sommerfeld factors are thus given by
\begin{equation}
  S_l^{(\gamma g, Z g)}=  S_l^{(\textbf{8})} \qquad S_l^{(\gamma \gamma, \gamma Z, ZZ, WW, Zh)}=  S_l^{(\textbf{1})}\,.
\label{eq:Sbos}
\end{equation}

Finally, we also consider the co-annihilation process $\psi \bar \psi \rightarrow \bar q q$. When the dark matter has a  tree-level Yukawa interaction with the quark $q$, the co-annihilation process is mediated by the t-channel exchange of a fermionic mediator, as well as by the s-channel exchange of a gluon. If this is not the case,  only the s-channel process is relevant. As a consequence, the $SU(3)_c$ representation of the initial state crucially depends on the concrete quark produced in the final state. Namely, if only the s-channel mediates the co-annihilation,  the initial state is a pure octet. On the other hand, if the s- and t-channels interfere, the initial state is combination of singlet and octet representations, therefore the total Sommerfeld factor can not be expressed in a simple form analogously to Eqs.~(\ref{eq:Sgg})-(\ref{eq:Sbos}); instead, we fully decompose the squared matrix element into the part corresponding to a singlet and an octet initial state, and multiply each contribution with $S_l^{(\textbf{1})}$ and $S_l^{(\textbf{8})}$, respectively.
\subsection{Results of the relic density calculation}
\label{sec:relic_density_results}
\begin{figure}[h!]
 \begin{center}     
   \begin{tabular}{cc}
   \hspace{-0.5cm}  \includegraphics[width=8.7cm]{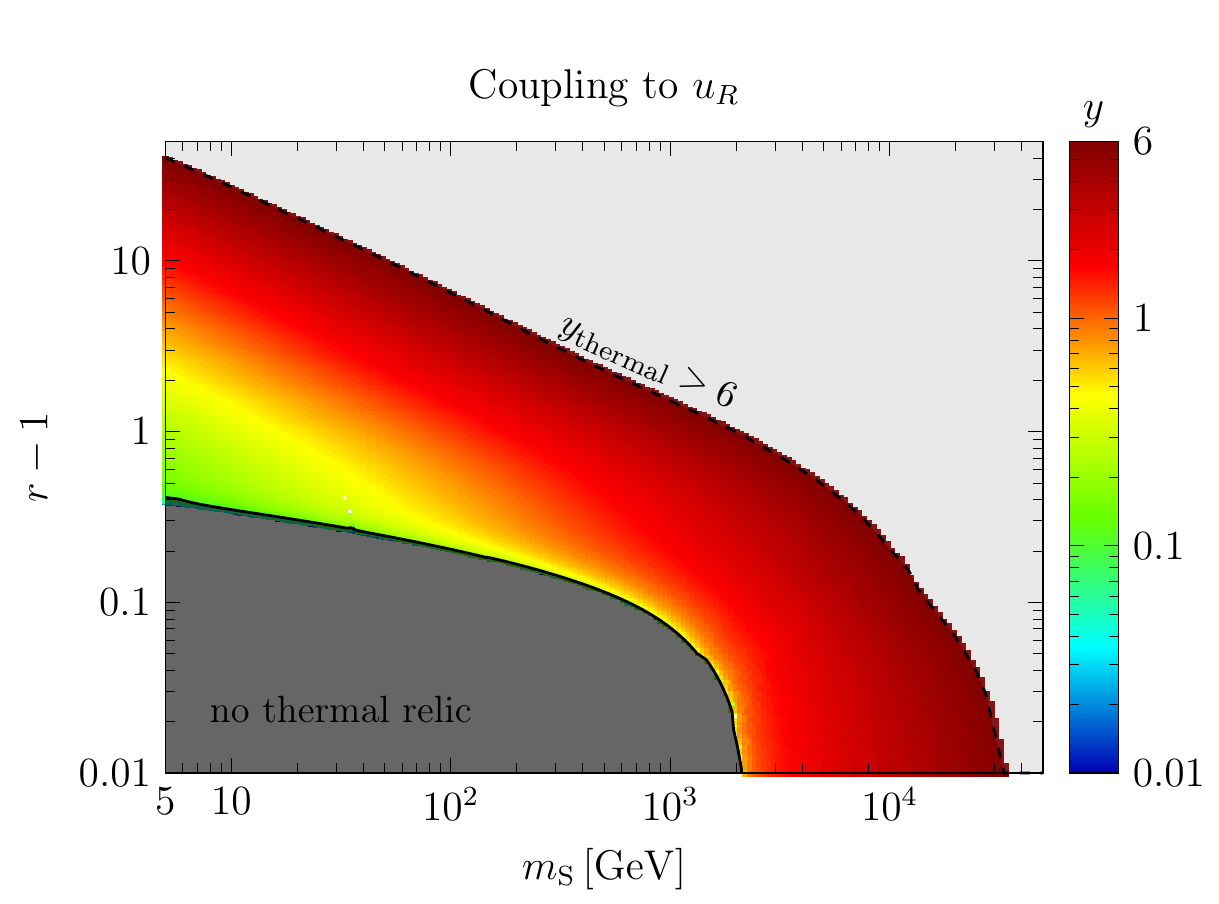}&  \includegraphics[width=8.7cm]{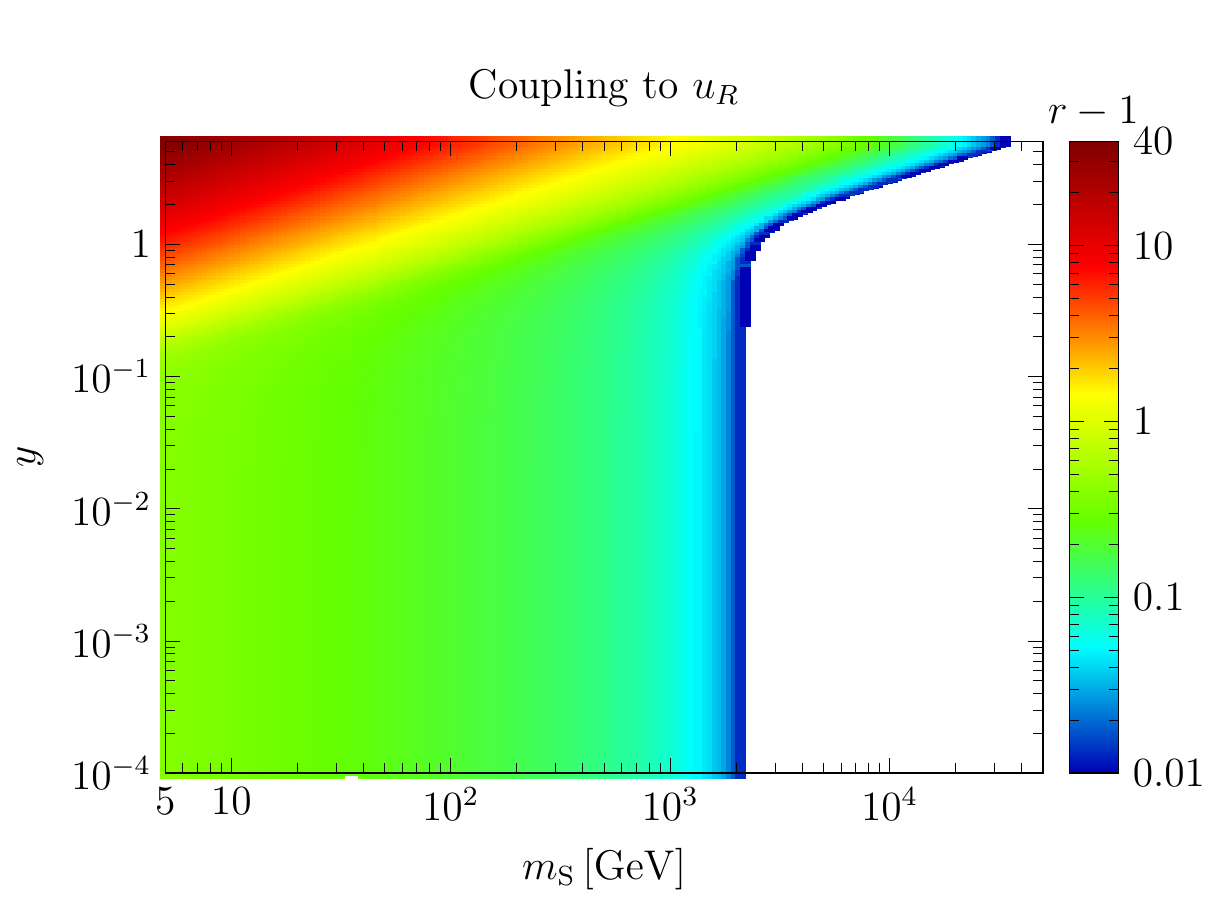} 
   \end{tabular}
 \end{center}
  \caption{\small Viable parameter space of the scalar dark matter
    model. In the
    left panel, the color gradient corresponds to values of the Yukawa
    coupling giving rise to the observed relic abundance,
    while in the right panel, the color gradient
    corresponds to values of $r-1$ (with $r=m_\psi/m_S$) leading to $\Omega_\text{DM} h^2
    \simeq 0.12$. Note that for definiteness, these plots are for the
    case of dark matter coupling to $u_R$, however the results for
    coupling to $d_R$ are practically identical (see discussion in the
    text).}
\label{fig:viab}
\end{figure}

In this work, we consider the case of dark matter coupling either to the right-handed up-quark $u_R$, or to the right-handed down-quark $d_R$. Both for the annihilation and the co-annihilation processes, the dominant contribution to the annihilation cross section arises either from the Yukawa interaction or from the QCD interaction, the latter being identical for coupling to $u_R$ and $d_R$. Hence, up to sub-percent corrections arising from diagrams involving electromagnetic or weak couplings, {the Yukawa coupling corresponding to the observed relic density} is identical in both of these scenarios. Note however that we fully include all sub-leading effects in our numerical calculations.

In Fig.~\ref{fig:viab} we show the viable parameter space in the plane
spanned by the dark matter mass $m_S$ and the relative mass difference $r-1$, as well as
in the plane spanned by $m_S$ and the Yukawa coupling $y$. The color
gradient corresponds to the value of the Yukawa coupling in the left
panel and to the value of $r-1$ in the right panel required to match
the observed relic abundance $\Omega_\text{DM} h^2 \simeq 0.12$.  In both
panels, the region at small $m_S$ and/or small
$r$, i.e.~the dark grey in the left panel, corresponds to
parameters for which the co-annihilation processes involving only gauge
interactions are sufficient to suppress the relic density below the
observed value. Besides, in the left panel, the light grey region at large $m_S$ and/or
$r$ corresponds to Yukawa couplings larger than $y=6$, which we choose
as perturbativity limit.\footnote{Our analysis includes the loop
  process $SS\rightarrow gg$, that depends on $ y^2
  g_s^2/16\pi^2$. Requiring $ y^2 g_s^2/16\pi^2 \lesssim 1$, such that the one-loop calculation becomes reliable, requires {$y \lesssim 6$}.}
Notice that for the most degenerate scenario discussed in this work,
$r=1.01$, the Sommerfeld corrections lead to a change in the relic
density of up to $\simeq 15 \%$. It is also worth mentioning that the
Sommerfeld effect can both increase as well as decrease the relic
density, depending on the dark matter mass and the mass
splitting. This is due to the different sign of the effective coupling
$\alpha$ depending on the initial color state of the annihilation
process (\emph{c.f.} Section~\ref{subsec:sommerfeld}).

\section{Direct detection constraints}
\label{sec:direct-detect-constr}
\begin{figure}
\begin{center}
\begin{tabular}{cc}
\raisebox{-.5\height}{\includegraphics[scale=0.65]{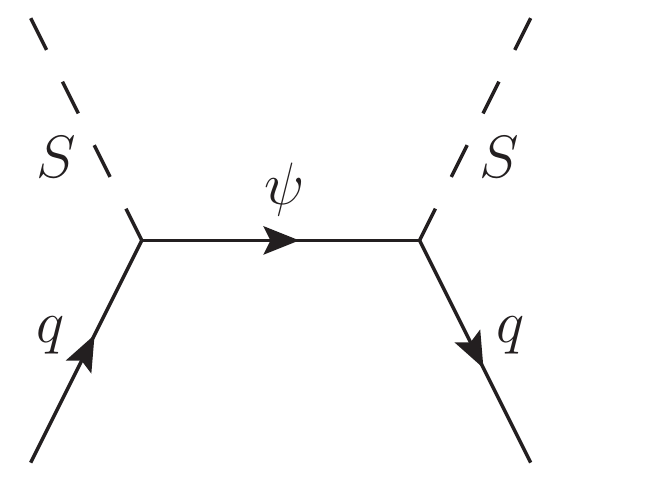}} & \raisebox{-.5\height}{\includegraphics[scale=0.65]{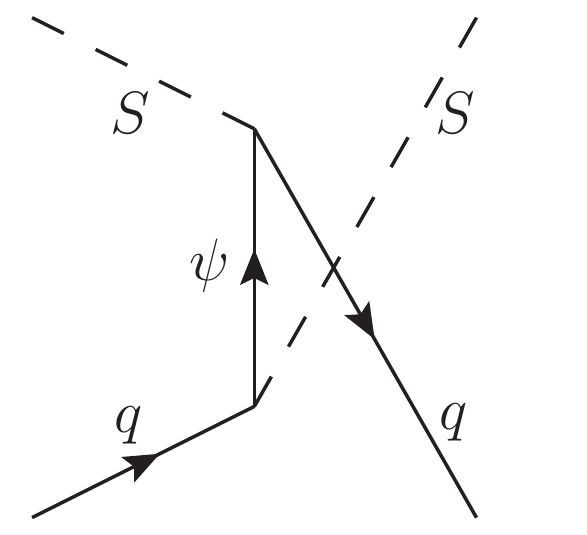}}
\end{tabular}\\[0.5cm]
\begin{tabular}{ccc}
\raisebox{-.5\height}{\includegraphics[scale=0.65]{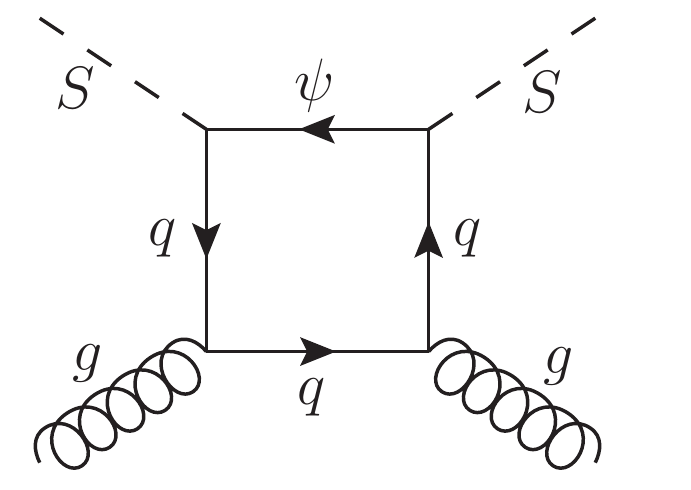}} & \raisebox{-.5\height}{\includegraphics[scale=0.65]{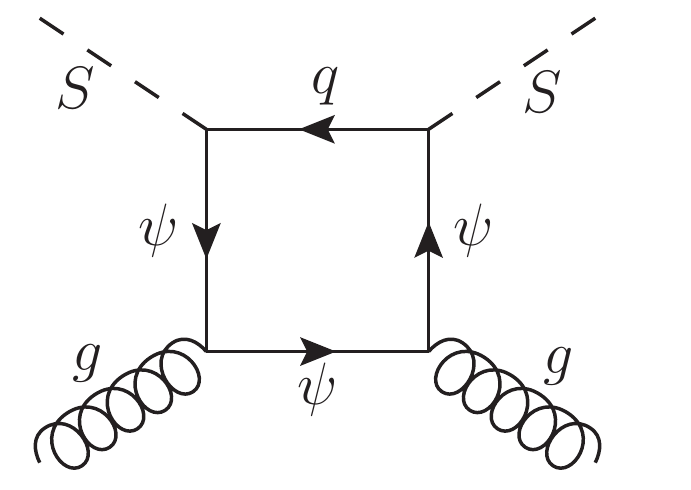}} & \raisebox{-.5\height}{\includegraphics[scale=0.65]{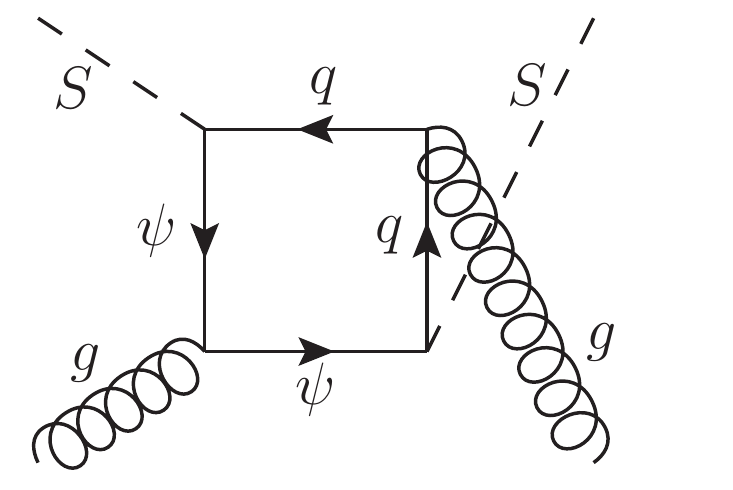}}
\end{tabular}
\end{center}
\caption{\small Diagrams contributing to the effective dark matter-nucleon coupling. Here, $q$ is short for either $u$ or $d$. Additional diagrams obtained by crossing initial or final states are not shown.}
\label{fig:DD_diagrams}
\end{figure}

The interaction of dark matter particles with nucleons leads to potentially observable signatures in direct detection experiments. In the model under study in this paper, this interaction is described by the Feynman diagrams shown in Fig.~\ref{fig:DD_diagrams}. The diagrams in the upper row correspond to the tree-level scattering off a light quark $q$, being in our case $q=u$ or $d$,  through the exchange of a fermionic mediator, while the one-loop diagrams depicted in the lower row lead to an effective interaction of the dark matter particle with gluons.

The effective interaction of $S$ with a light quark $q$ can be cast as the sum of a scalar and a twist-2 contribution~\cite{Drees:1993bu}:
\begin{align}
 \mathcal{L}_{q} &= C_S^q \, m_q S^2 \bar{q} q \, \, + \, \,C_T^q\, (\partial_\mu S) (\partial_\nu S) \mathcal{O}_{q,\text{twist-2}}^{\mu \nu} \, \,, \label{eq:Lq_eff} \\
 \text{with } \, \, \mathcal{O}_{q,\text{twist-2}}^{\mu \nu} &\equiv\frac{i}{2} \left( \bar q \gamma^\mu \partial^\nu q + \bar{q} \gamma^\nu \partial^\mu q - \frac{g^{\mu \nu}}{2} \bar q \slashed{\partial} q \right) \,. \label{eq:OTwist2}
\end{align}
By integrating out the mediator $\psi$ in the diagrams shown in the upper row of Fig.~\ref{fig:DD_diagrams}, and matching the result to the effective Lagrangian in Eq.~(\ref{eq:Lq_eff}), we obtain
\begin{align}
C_S^q = \frac{y^2}{4 m_S^2} \frac{2 r^2-1}{(r^2-1)^2} \, \, , \, \, C_T^q = \frac{y^2}{m_S^4} \frac{1}{(r^2-1)^2} \,,
\label{eq:CSq_CTq}
\end{align}
where $q=u$ ($q=d$) for dark matter coupling to $u_R$ ($d_R$). 

On the other hand, the diagrams in the lower row of Fig.~\ref{fig:DD_diagrams} induce an effective coupling of dark matter to gluons, which has been recently discussed in~\cite{Hisano:2015bma}. Similarly to the well-studied simplified model of Majorana dark matter with a colored scalar mediator~\cite{Hisano:2010ct}, the scattering amplitude can be decomposed into a short-distance and a long-distance contribution. This separation is made according to the momentum scale dominating the loop integration: the part of the amplitude arising from loop-momenta of the order of the mass scale of the heavy particles ({\it i.e.} the dark matter particle or the fermionic mediator) leads to the short-distance contribution, while the part arising from loop-momenta of the order of the light quark masses, leads to the long-distance contribution. The latter involves values of the strong coupling constant at a non-perturbatively small momentum scale, and hence can not be reliably calculated in perturbation theory. However, this contribution is implicitly contained in the parton distribution function of the light quarks in the nucleon, and hence only the short-distance contribution must be taken into account in the computation of  the one-loop diagrams. Defining the effective Lagrangian for the dark matter-gluon interaction as
\begin{align}
 \mathcal{L}_{g} = C_S^g \, \frac{\alpha_S}{\pi} S^2 \, G^{\mu \nu} G_{\mu \nu} \,, \label{eq:Lg_eff} 
\end{align}
the short-distance contribution to $C_S^g$ reads, assuming the limit $m_q\ll m_\psi-m_S$
~\cite{Hisano:2015bma},
\begin{align}
 C_S^g = \frac{y^4}{24 m_S^2} \frac{1}{r^2-1} \,.
 \label{eq:CSg}
\end{align}

\begin{figure}
\begin{center}     
   \begin{tabular}{cc}
   \hspace{-1.3cm}
     \includegraphics[scale=0.55]{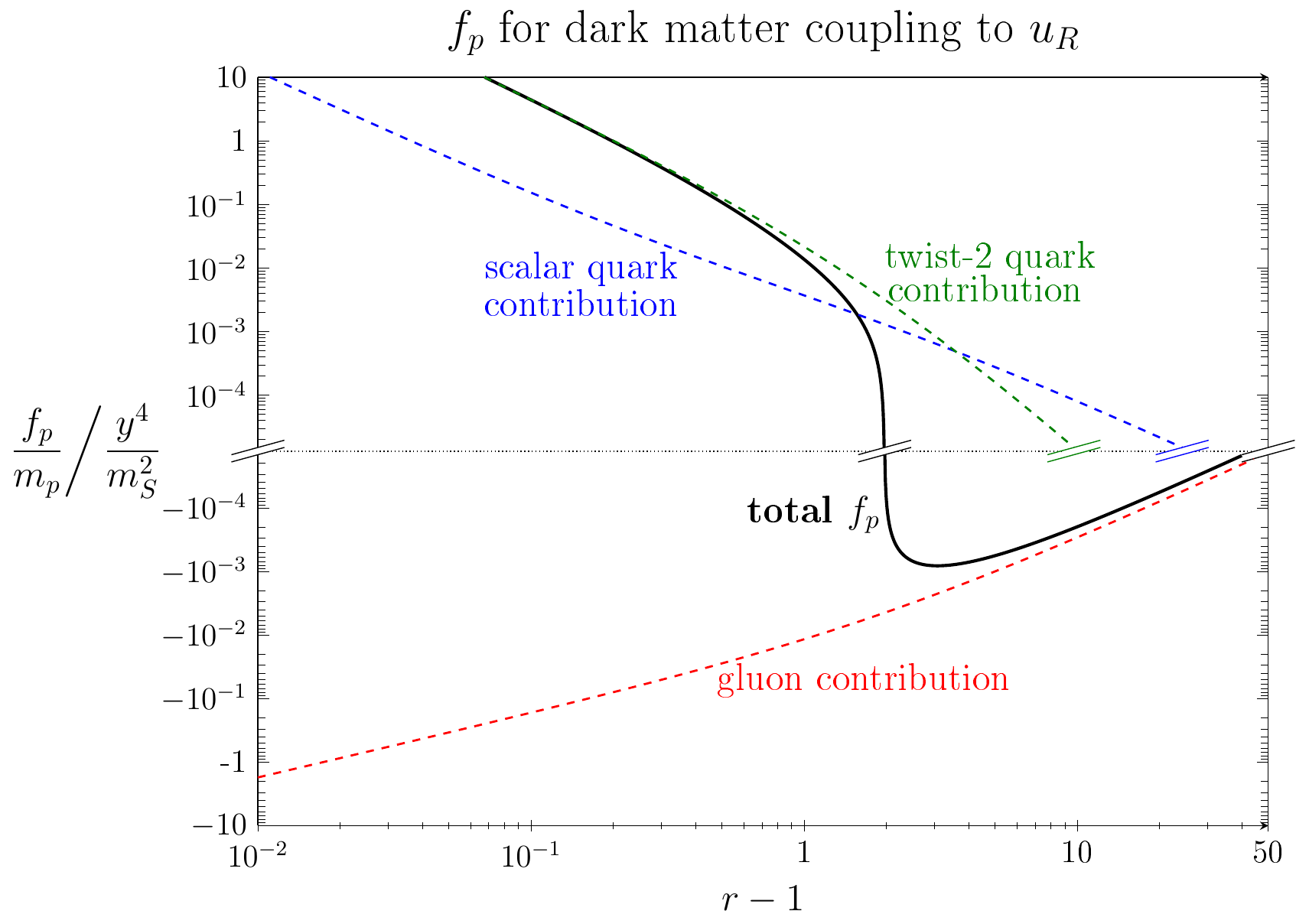}&
     \hspace{0.3cm}
     \includegraphics[scale=0.58208]{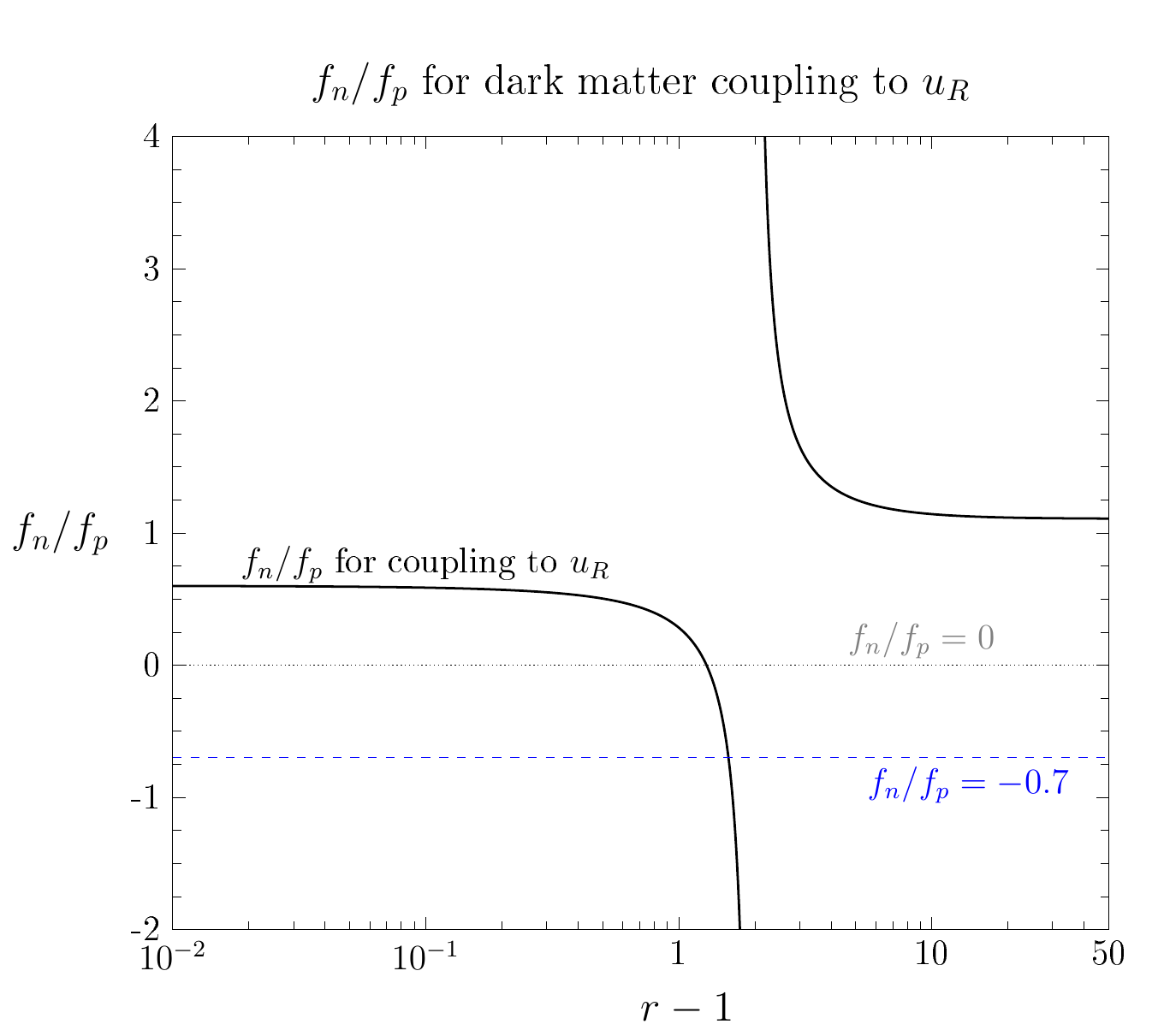}
   \end{tabular}
 \end{center}
  \caption{\small Left panel: dark matter-proton coupling $f_p/m_p$, in units of $y^4/m_S^2$, for coupling to $u_R$. Right panel: ratio of the neutron-to-proton coupling of dark matter, also for coupling to $u_R$.}
  \label{fig:fpfn_plots}
  \end{figure}

From the effective Lagrangians to the partons, Eqs.~(\ref{eq:Lq_eff}) and~(\ref{eq:Lg_eff}), one can calculate the effective spin-independent coupling of the dark matter particle $S$ to a nucleon $N$, the result being~\cite{Drees:1993bu}
\begin{align}
\frac{f_N}{m_N} = C_S^q f_{T_q}^{(N)} + \frac34 C_T^q m_S^2 (q^{(N)}(2) + \bar{q}^{(N)}(2)) - \frac89 C_S^g f_{T_G}^{(N)} \,,
\label{eq:fN}
\end{align}
where $f_{T_q}^{(N)}$, $f_{T_G}^{(N)}$ are mass fractions and $q^{(N)}(2), \bar{q}^{(N)}(2)$ are the second moments of the parton distribution functions; for our numerical analysis we use the values in~\cite{Hisano:2015bma}. We show in Fig.~\ref{fig:fpfn_plots}, left panel, the effective dark matter-proton coupling $f_p/m_p$, expressed in units of $y^4/m_S^2$ (which is the common pre-factor to each of the terms in Eq.~(\ref{eq:fN})), as a function of $r-1$, for the case of dark matter coupling to $u_R$. As apparent from the plot, the coupling of dark matter to gluons, corresponding to the last term in Eq.~(\ref{eq:fN}), interferes destructively with the contributions arising from the dark matter scattering off quarks, leading to a vanishing dark matter-proton coupling at $r \simeq 3.0$. A similar behavior is found for $f_n$, although the total destructive interference occurs at a slightly different mass splitting $r \simeq 2.3$. 

Finally, from the effective dark matter-nucleon
couplings $f_p$ and $f_n$, one can calculate the total
spin-independent cross section for dark matter scattering off a
nucleus with charge $Z$ and mass number $A$. At zero momentum
transfer, which is the regime relevant for direct detection
experiments, it is given by the coherent sum of $f_p$ and $f_n$:
\begin{align}
\sigma_A = \frac{m_A^2}{\pi (m_S+m_A)^2} \left[ Z f_p + (A-Z)f_n \right]^2 \,.
\label{eq:sigma_general}
\end{align}
We find that, among all current experiments, the LUX experiment~\cite{Akerib:2013tjd} gives, in most of the parameter space, the best limits to the simplified model under scrutiny in this paper~\footnote{After this work has been completed, the LUX collaboration presented a reanalysis of the 2013 data, leading to improved upper limits on the dark matter scattering cross section~\cite{Akerib:2015rjg}. However, none of our conclusions would change qualitatively when including the updated limits.}. LUX is based on a xenon target and, as is well known, has least sensitivity when $f_n/f_p \simeq -0.7$, the precise value being dependent on the xenon isotope considered. In this case, commonly referred to as ``maximal isospin violation'',  the total scattering cross section vanishes. We show in the right panel of Fig.~\ref{fig:fpfn_plots} the ratio $f_n/f_p$ as a function of the mass splitting, for the case of dark matter coupling to $u_R$. As follows from the plot, in this scenario maximal isospin violation occurs at $r \simeq 2.6$ for dark matter coupling to $u_R$ (this occurs at $r \simeq 3.3$ for dark matter coupling to $d_R$). Hence, we fully take into account the actual value of $f_n/f_p$ predicted by the model by defining the \emph{effective dark matter-proton cross section} $\sigma_p^{\text{eff}}$ via
\begin{align}
  \sigma_{p}^{\text{eff}}= \sigma_{p} \, \cdot \, \frac{\sum_{i\in {\rm isotopes}} \xi_i \left(Z+(A_i-Z) f_n/f_p\right)^2 }{\sum_{i\in {\rm isotopes}} \xi_i A_i^2 }\, ,\label{eq:speff}  
\end{align}
where $\xi_i$ is the natural relative abundance of the xenon isotope $i$. In that way, $\sigma_{p}^{\text{eff}}$ can be interpreted as the dark matter-proton scattering cross section that under the assumption of $f_p=f_n$ would lead to the same number of events in LUX as the actual dark matter-proton cross section $\sigma_p$, when taking into account the ratio $f_n/f_p$ predicted by the model. Consequently, $\sigma_{p}^{\text{eff}}$ is the quantity that can be compared to the upper limit derived by the LUX collaboration~\cite{Akerib:2013tjd}, which was derived under the assumption that $f_p=f_n$.

\begin{figure}[h!] 
\begin{tabular}{cc}    
   \hspace{-0.9cm} \includegraphics[width=8.9cm]{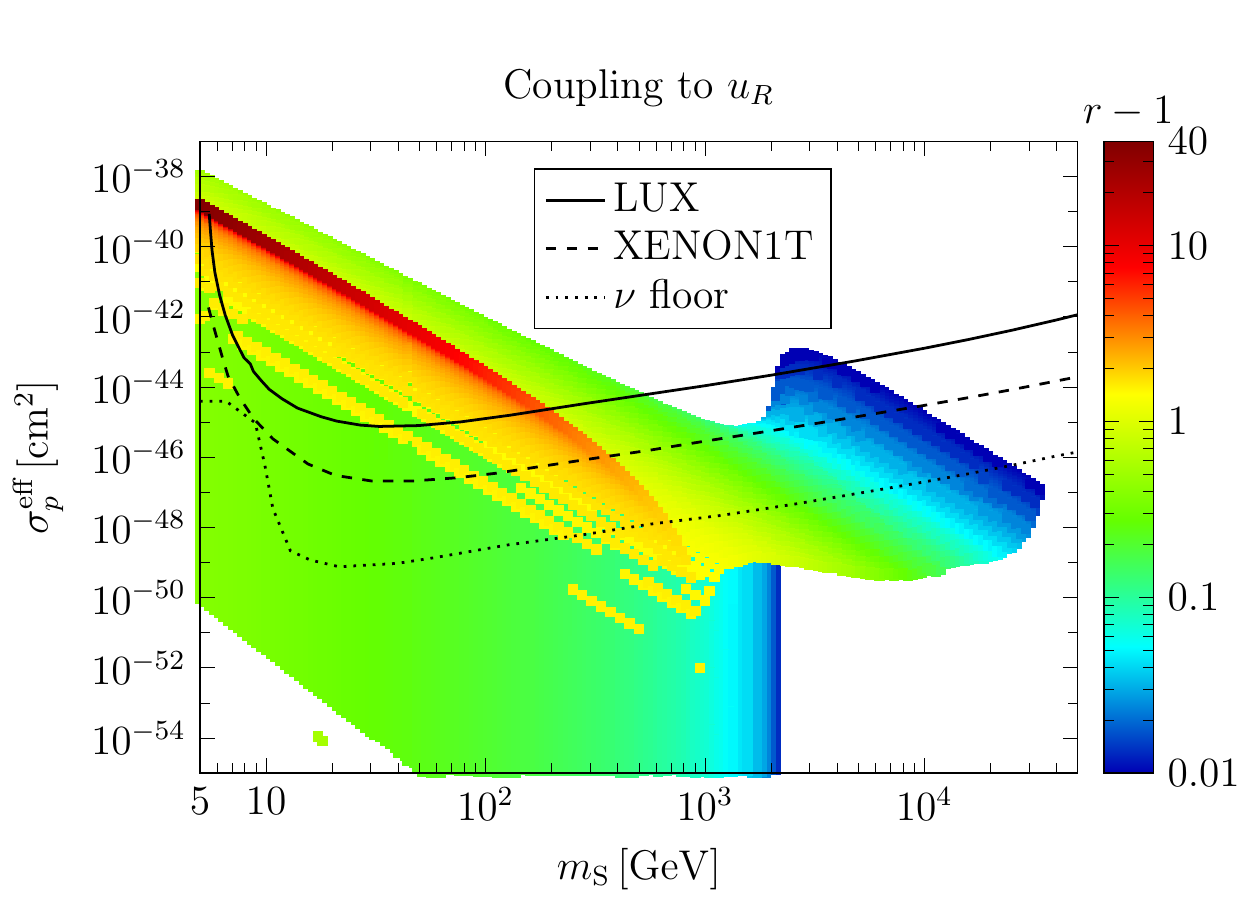}&
   \includegraphics[width=9.35cm]{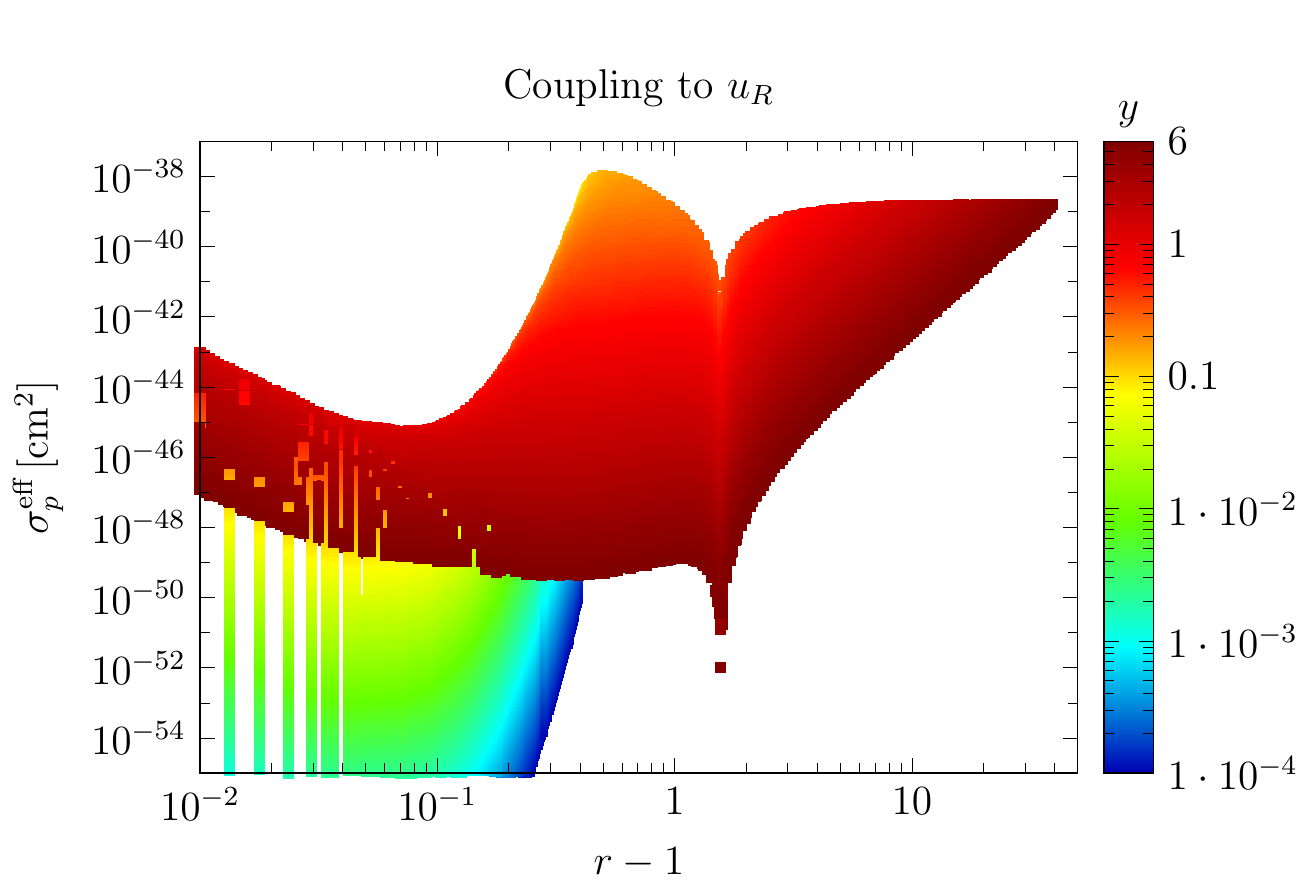}
  \end{tabular}
  \caption{\small Effective dark matter-proton scattering cross-section $\sigma_p^\text{eff}$, assuming that the thermal production of $S$ accounts for all of the observed dark matter, as a function of the dark matter mass (left panel) and as a function of the ratio between the fermion mediator, $\psi$, and the dark matter, $S$ (right panel). The color gradient in the left plot corresponds to different values of $r=m_\psi/m_S$, while in the right plot, to different values of the Yukawa coupling $y$. The solid, dashed and dotted lines in the left plot show, respectively, the present upper limit on the effective cross section from the LUX experiment, the projected sensitivity of the XENON1T experiment, and the neutrino floor calculated in~\cite{Billard:2013qya}. 
}
  \label{fig:sSp}
\end{figure}
 
In Fig.~\ref{fig:sSp}, we show the effective dark matter-proton scattering cross section $\sigma_{p}^\text{eff}$ in two different projections of the thermal parameter space. In the left panel, the color gradient corresponds to different values of the relative mass difference between the vector-like mediator and the dark matter particle. Let us emphasize that for every dark matter mass and mass splitting, we fix the Yukawa coupling $y$ by the requirement of matching the relic abundance, \textit{cf.}~section~\ref{sec:relic_density_results}. As it can be seen from the plot, there is a huge scatter in the possible values of the scattering cross section over many orders of magnitude. Generally speaking, for a fixed dark matter mass, the scattering cross section gets larger for more and more degenerate scenarios, {\it i.e.}~smaller $r$, due to the resonant enhancement of the dark matter-nucleon coupling $f_N$ in the limit $r\rightarrow 1$, which follows from Eqs.~(\ref{eq:CSq_CTq}), (\ref{eq:CSg}) and (\ref{eq:fN}). However, as it can be inferred from Fig.~\ref{fig:viab}, for a given mass splitting $r$, co-annihilations imply a lower limit on the dark matter mass from the requirement of matching the observed relic density. Hence, points in the left panel of Fig.~\ref{fig:sSp} corresponding to a fixed, small value of $r-1$ show a turnaround at this dark matter mass, which is clearly visible in the plot e.g.~for $r=1.01$ (dark blue points). Furthermore, we observe that for a given value of $m_S \lesssim 200$ GeV, the effective scattering cross section converges to a fixed value in the limit of $r \gg 1$. This can easily be understood by noting that in this limit, the most relevant process for the freeze-out of dark matter is the one-loop annihilation of dark matter into gluons, scaling as $1/r^4$, which is precisely the same asymptotic behaviour of the scattering cross section for $r \gg 1$, as it can be seen from Eqs.~(\ref{eq:CSq_CTq}), (\ref{eq:CSg}), (\ref{eq:fN}) and (\ref{eq:sigma_general}).
 
In the left panel of Fig.~\ref{fig:sSp}, we also show the current upper limit from LUX, as well as the projected upper limit for the XENON1T experiment, taken from~\cite{Aprile:2012zx}. In addition, we also show in the plot the ultimate reach of (non-directional) direct detection experiments, given by the scattering cross section corresponding to coherent neutrino scattering in the detector~\cite{Billard:2013qya}. As apparent from the plot, for $m_S \lesssim 200-300$ GeV, large parts of the parameter space are already excluded by LUX, and for the most degenerate scenarios even dark matter masses around 2 TeV are already ruled out. In the near future, XENON1T will continue closing in the region of the parameter space of thermal dark matter, in particular for dark masses above the TeV scale, which will be difficult to probe at the LHC (see the discussion in the next chapter).
 
Finally, in the right panel of Fig.~\ref{fig:sSp} we show the effective dark matter-proton scattering cross section as a function of $r-1$, with the color coding corresponding to the Yukawa coupling $y$, and fixing the dark matter mass by the relic density requirement. In this projection of the parameter space, the suppression of the effective cross section at $r\simeq 2.6$ becomes apparent, due to the maximal isospin violation occurring at this value of the mass ratio.

\section{Constraints from searches at the LHC}
\label{sec:collider-constr}

The model discussed in this work can be efficiently probed at the LHC through the production of the fermionic mediator $\psi$. After being produced, the mediator decays promptly into the dark matter particle $S$ and a light quark; hence, the signature of interest consists of (at least) two jets plus missing transverse $E_T$. In the following, we first discuss the relevant contributions to the production cross section of the fermionic mediator $\psi$, and subsequently we derive constraints on the model by reinterpreting a recent ATLAS search for multiple jets plus missing $E_T$. 

\subsection{Production of mediator pairs}
\label{sec:production_mediator_pairs}

\begin{figure}
\begin{center}
\begin{tabular}{cccc}
$\mathcal{M} \propto g_s^2$: & \raisebox{-.5\height}{\includegraphics[scale=0.65]{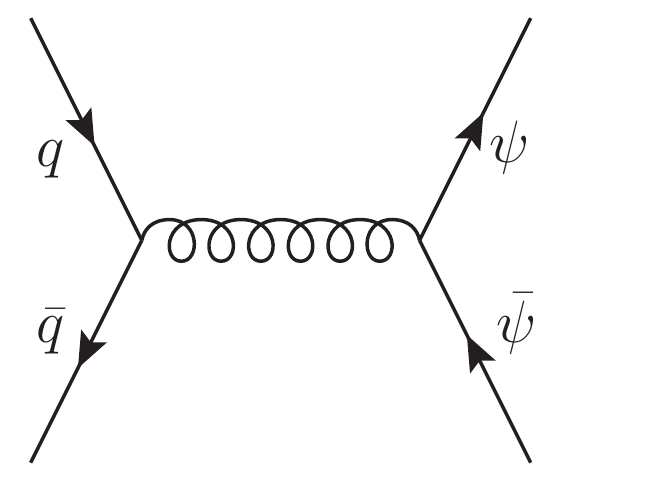}} & \raisebox{-.5\height}{\includegraphics[scale=0.65]{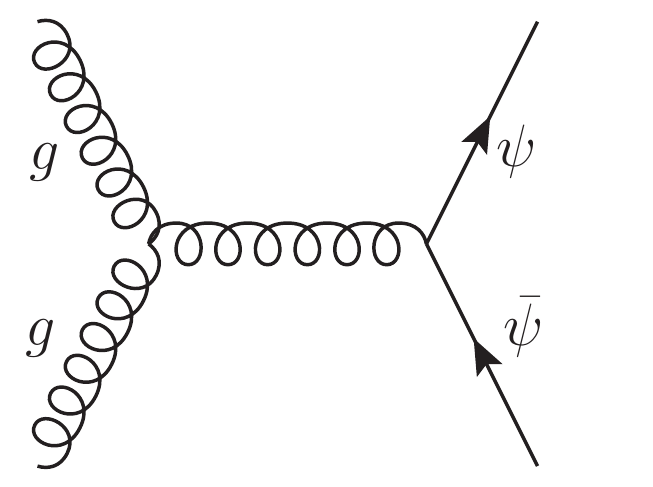}} & \raisebox{-.5\height}{\includegraphics[scale=0.65]{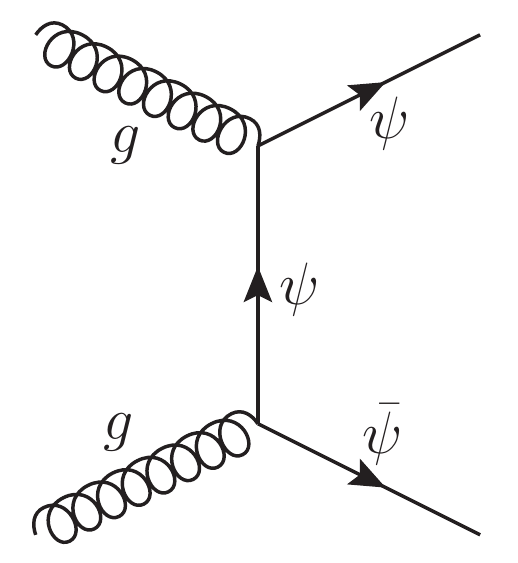}}
\end{tabular}
\begin{tabular}{ccc}
\hspace{-4.8cm}
$\mathcal{M} \propto y^2$: & \raisebox{-.5\height}{\includegraphics[scale=0.65]{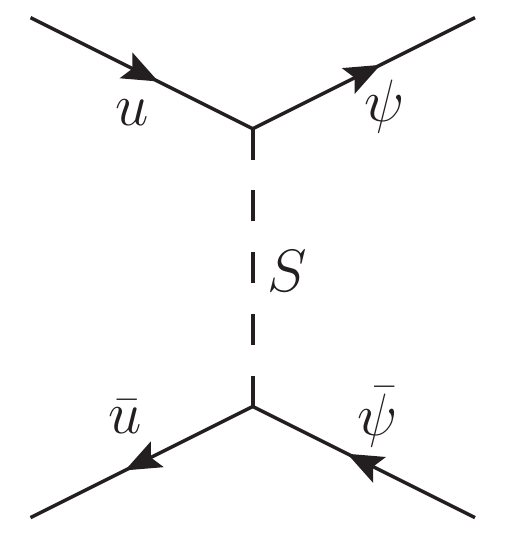}} & \hspace{1cm}\raisebox{-.5\height}{\includegraphics[scale=0.65]{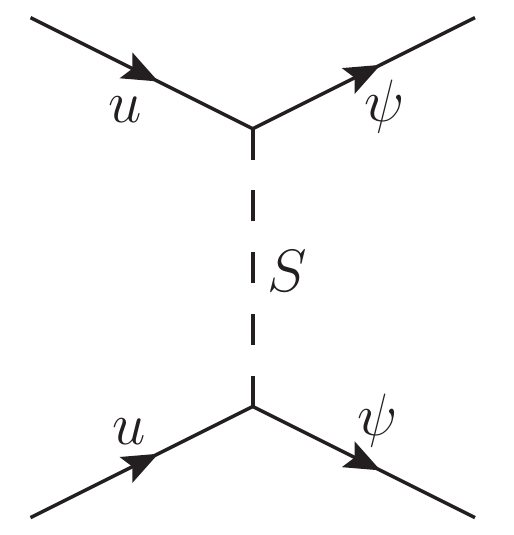}}
\end{tabular}
\end{center}
\caption{\small Feynman diagrams of the relevant production modes of mediator pairs in proton-proton collisions. Additional diagrams obtained from crossing or charge conjugation of the initial and final state are not shown.}
\label{fig:lhc_diagrams}
\end{figure}

We consider the production of mediator pairs $\psi \psi$, $\psi \bar
\psi$, and $\bar \psi \bar \psi$ in proton-proton collisions at a
center-of-mass energy $\sqrt{s} = 8$ TeV.\footnote{In
  addition to the processes $pp\rightarrow X X $, with $X=\psi, \bar
  \psi$ considered here, the production of mediators through the
  process $pp\rightarrow X S $ provide extra contributions to the
  multiple jets plus missing $E_T$ analysis. Assuming $y=y_{\text{thermal}}$,
  we have checked that the total production cross sections satisfy
  $\sigma (pp\rightarrow X S)< \sigma (pp\rightarrow X X)$ and that
  the enhancement in the production cross-section obtained when
  including $\sigma (pp\rightarrow X S)$ does not affect significantly
  the limits that are presented in
  Fig.~\ref{fig:results_thermal_plane}.  }. The relevant Feynman
diagrams, shown in Fig.~\ref{fig:lhc_diagrams}, can be divided into
two categories: first, a $\psi \bar \psi$ pair can be produced from a
$q \bar q$ or $gg$ initial state by QCD interactions, as shown in the
upper row of Fig.~\ref{fig:lhc_diagrams}. Secondly, {for the case of
  dark matter coupling to $u_R$,} a pair of $\psi \psi$, $\psi \bar
\psi$ or $\bar \psi \bar \psi$ can be produced from a $uu$, $u \bar
u$, or $\bar u \bar u$ initial state, respectively, through the
$t$-channel exchange of the dark matter particle $S$, as shown in the
lower row of Fig.~\ref{fig:lhc_diagrams}. {An analogous statement
  holds for dark matter coupling to $d_R$, by exchanging $u$ with $d$
  in the initial states.} The amplitudes corresponding to the QCD
induced diagrams are independent of the dark matter mass and the
Yukawa coupling $y$, in contrast to the processes arising from the
Yukawa interaction.

In the model discussed in this work, the requirement of correctly
reproducing the observed dark matter abundance via thermal freeze-out
 can involve rather large Yukawa couplings, as
  can be seen in e.g.~Fig.~\ref{fig:viab}. On the other hand, the
    strong coupling $g_s \simeq 1$, therefore the production of
  mediator pairs can be dominated either by the processes
  involving the Yukawa coupling or by the strong
    coupling. Furthermore, for dark matter coupling to $u_R$
    ($d_R$), the production channel $uu \rightarrow \psi \psi$ {($dd
    \rightarrow \psi \psi$)}, which is driven solely by the Yukawa
  coupling, can have a larger cross section compared to processes
  involving a $q \bar{q}$ or $gg$ pair in the initial state, due to
  the large parton distribution function of the up-quark
  {(down-quark)} in the proton. In our analysis, we calculate the
  leading-order cross section for the final states $\psi \psi, \psi
  \bar \psi, \bar \psi \bar \psi$ using {\tt
    CalcHEP}~\cite{Belyaev:2012qa}, selecting the {\tt cteq6l} parton
  distribution function. In order to include next-to-leading order
  corrections, we parametrize the total cross section as
\begin{equation}
\sigma^{\rm full}= K \, \sigma^{\rm LO} \,,
  \label{eq:sigNLO}
\end{equation}
The $K$-factor has not been fully calculated for this model, consequently we treat $K$ as a source of uncertainty, varying it in the range $[0.5, 2]$.~\footnote{For comparison,~the~$K$-factor for~$t \bar t$~production,~which is analogous to the QCD production of~$\psi \bar \psi$,~is~$1.7$~\cite{Kidonakis:2010dk}.}

\begin{figure}
\begin{center}
\hspace{-1.8cm}
\includegraphics[scale=1.25]{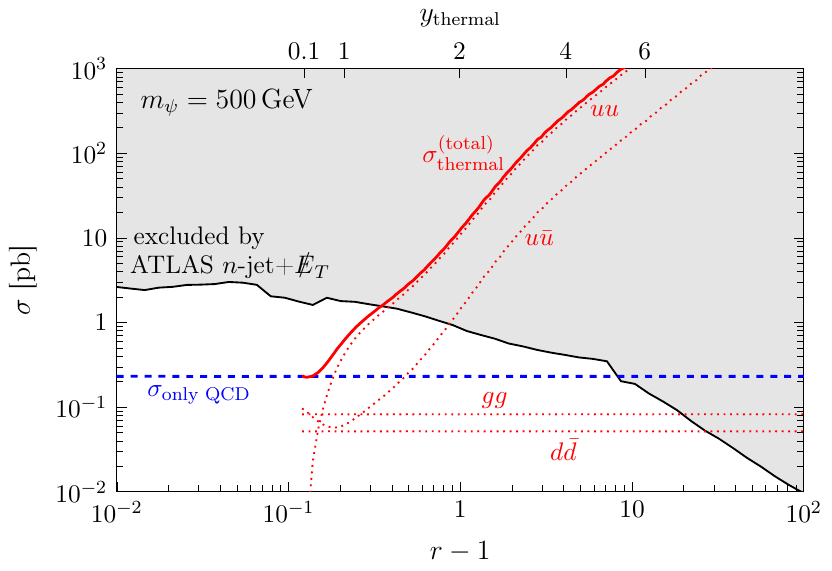}
\end{center}
\caption{\small Total cross section for $pp \rightarrow \psi \psi, \psi \bar \psi, \bar \psi \bar \psi$ at $\sqrt{s} = 8$ TeV for $m_\psi = 500$ GeV, together with the individual contributions from different initial states. For the red curves, the Yukawa coupling $y$ is set to the value corresponding to $\Omega_\text{DM} h^2 = 0.12$, while the blue dashed curve indicates the production cross section only from pure QCD processes.}
\label{fig:sigma_prod_excl}
\end{figure}

We show in Fig.~\ref{fig:sigma_prod_excl} the production cross section
for the various initial states for the exemplary choice {of dark
  matter coupling to $u_R$, and choosing $m_\psi = 500$ GeV as well as
  $K=1$}. The red curves were calculated for the value of the Yukawa
coupling that leads to the observed dark matter abundance via thermal
freeze-out, $y = y_\text{thermal}$, while the blue dashed curve shows,
for comparison, the predicted cross section when taking into account
only pure QCD processes in the production of mediator
pairs. Notice the rise of
the total cross section for large values of the mass ratio, that is entirely due to the Yukawa induced
processes. In particular, for $m_\psi/m_S \gtrsim 1.2$, the production
channel $u u \rightarrow \psi \psi$ dominates, due to the large
thermal Yukawa coupling as well as due to the large parton distribution
function of the up-quark in the proton, as discussed above.
From Fig.~\ref{fig:sigma_prod_excl}, it also follows that the cross section
  for the process $u \bar u \rightarrow \psi \bar \psi$ is lower than 
  the pure QCD cross section in the range $1.1 \lesssim m_\psi/m_S
  \lesssim 1.3$. This is due to the destructive interference of the QCD
  and Yukawa driven contributions to the process.

\subsection{Constraints derived from the multijet ATLAS analysis}

The production of $\psi$ can be detected at the LHC through its decay
into a quark and the dark matter particle $S$. Ignoring the 
higher-order corrections, the event topology always consists
of two jets arising from the decay of the two mediators, together with
missing $E_T$ associated to the dark matter particles 
(this process was considered in~\cite{Chang:2013oia}, neglecting, however,
the impact of co-annihilations and higher-order effects in the calculation
of the dark matter abundance). On the other hand, the 
radiation of quarks or gluons from the initial, final or an
intermediate state can lead to additional jets which could also  be detected.
 Considering these extra jets from higher-order processes
is important due to two reasons: first, if the absolute mass splitting
$m_\psi - m_S$ is below $\sim 50-100~\text{GeV}$, the
jet arising from the decay of $\psi$ is too soft to be detected 
at ATLAS or CMS. In that case, the additional emission of
one or more hard jets is necessary for the detection of the
event (for an analysis of the constraints of the model employing 
monojet searches, see~\cite{Papucci:2014iwa,Chang:2013oia}).
Secondly, even if the mass splitting of the dark matter
particle and the mediator is sufficiently large, the
signal-to-background ratio of the $n$-jet $+\slashed{E}_T$ signal can
be larger for $n>2$ compared to $n=2$, depending on the
characteristics of the relevant Standard Model background
processes. Consequently, even if the absolute cross section is
smaller, the constraints derived from the multijet processes can be
more stringent than those obtained from the lowest order two-jet
topology.

In order to confront our model to available LHC data, we make use of the ATLAS search {\tt ATLAS-CONF-2013-047}~\cite{TheATLAScollaboration:2013fha} for 2-6 jets plus missing $E_T$, at $\sqrt{s}=8$ TeV with a luminosity of $20.3 \text{ fb}^{-1}$. For the search, ten different signal regions were defined by the collaboration, differing mainly by the number of jets and the minimal amount of missing $E_T$. The dominant Standard Model processes giving rise to the relevant event topologies are the production of jets together with a weak gauge boson (which can decay invisibly, leading to the missing $E_T$ signature), as well as the production of top quarks. The expected size of the background in each signal region has been estimated in~\cite{TheATLAScollaboration:2013fha}, employing the commonly used approach of extrapolating the measured number of events from separate control regions, and additionally validating the method by directly using Monte Carlo generators for an estimation of the expected background rates. The ATLAS collaboration found no significant excess in any of the signal regions, and the corresponding observed (expected) $95 \%$ C.L. upper limits on the number of signal events in each signal region, which can be found in~\cite{TheATLAScollaboration:2013fha},  are denoted as $S_\text{obs}^{95}$ ($S_\text{exp}^{95}$).

On the other hand, for a given signal region $i$, the number of expected signal events can be cast as
\begin{align}
S_i = \sigma \cdot \epsilon_i \cdot \mathcal{L} \,,
\end{align}
where $\sigma$ is the total production cross section for pairs of $\psi$ as discussed in section~\ref{sec:production_mediator_pairs}, and $\mathcal{L} = 20.3 \text{ fb}^{-1}$ is the luminosity. Moreover, the efficiency $\epsilon_i$ is the probability that a given event passes all the selection cuts corresponding to the signal region $i$. This quantity depends on the topology of the involved Feynman diagrams, and hence has to be recalculated for a given model. For that purpose, we have implemented our model in {\tt FeynRules}~\cite{Alloul:2013bka}, and use {\tt MadGraph}~\cite{Alwall:2014hca} in order to simulate events for the production of mediator pairs with up to two additional jets. The partonic events are fed to {\tt PYTHIA 6}~\cite{Sjostrand:2006za}, which simulates the showering and hadronization process. As usual in this context, in that step one has to take care of the possible double counting in the additional jets, which should either be included at the matrix element level in the partonic event, \emph{or} as part of the showering process. In order to ensure a smooth transition between the two regimes, we employ the MLM matching scheme, using the parameters {\tt Qcut} $=m_\psi/4$ and {\tt SHOWERKT=T}. We explicitly checked that with this choice the differential jet distributions are smooth, as required for a physical meaningful matching scheme.

Finally, in order to apply the cuts corresponding to the ATLAS search, we use {\tt CheckMATE}~\cite{Drees:2013wra}, which also implements a detector simulation. Concretely, for given values of $m_S$, $m_\psi$, and the corresponding Yukawa coupling $y = y_\text{thermal}(m_S,m_\psi)$, we simulate $N_\text{ev}$ events as described above, and for every signal region $i$ {\tt CheckMATE} determines the number of events $N_\text{after cuts}^{(i)}$ passing all the cuts. Then, the efficiency can be obtained through $\epsilon_i = N_\text{after cuts}^{(i)} / N_\text{ev}$. As only a finite number of events can be generated, the efficiency obtained in this way is subject to Monte Carlo uncertainties. In order to take these into account in a conservative way, we replace $S_i$ by the $95 \%$ C.L. lower limit on $S_i$, given by $S_i - 1.96 \Delta S_i$, where $\Delta S_i$ is the $1 \sigma$ statistical error on the number of expected signal events\footnote{We generate up to $5 \cdot 10^6$ events per point in the parameter space in order to ensure that $\Delta S_i/S_i \lesssim 0.1-0.2$.}. Then, the $95 \%$ C.L. upper limit on the production cross section following from the observed (expected) number of events in the signal region $i$ is given by
\begin{align}
\sigma^{(95)}_{\text{obs},i} = \sigma \cdot \frac{S^{95}_{\text{obs},i}}{S_i - 1.96 \Delta S_i} \quad \,, \quad \sigma^{(95)}_{\text{exp},i} = \sigma \cdot \frac{S^{95}_{\text{exp},i}}{S_i - 1.96 \Delta S_i} \,.
\end{align}
Depending on the specific point in the parameter space, a different signal region provides the most stringent constraint. In order not to bias the final result by making a choice of the optimal signal region based on an over- or under-fluctuation, we choose for a given $m_S$ and $m_\psi$ the signal region giving rise to the smallest \emph{expected} upper limit $\sigma^{(95)}_{\text{exp},i}$, and then use the observed upper limit $\sigma^{(95)}_{\text{obs},i}$ in that signal region as our final upper limit on the production cross section. As an example, this upper limit is shown in Fig.~\ref{fig:sigma_prod_excl} for $m_\psi = 500$ GeV, together with the production cross section predicted by the model.

\begin{figure}[h!]
\begin{center}
\includegraphics[scale=1.5]{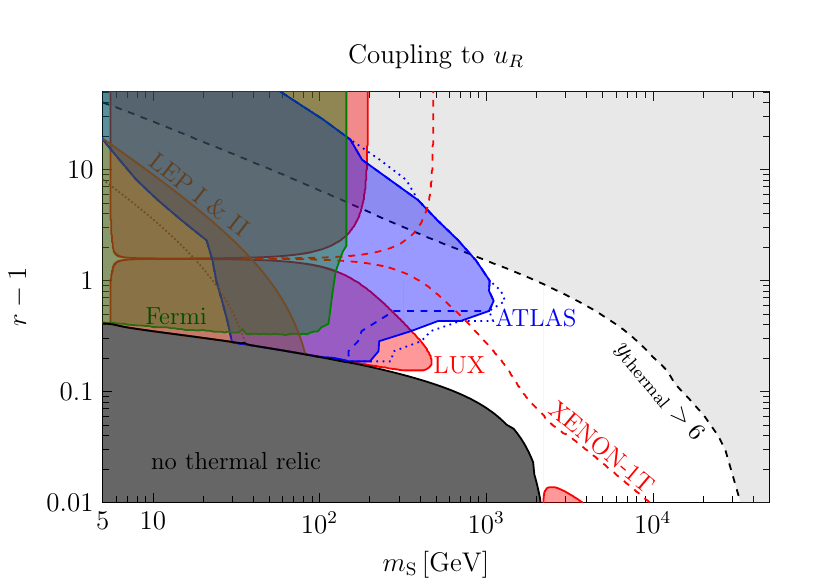}
\vspace{0.3cm}
\includegraphics[scale=1.5]{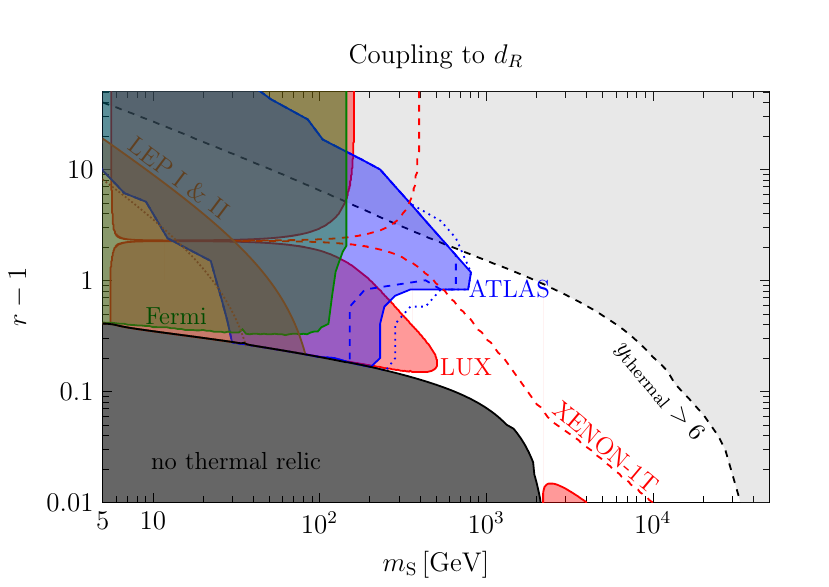}
\end{center}
\caption{\small Summary of constraints from direct and indirect detection and from collider searches projected on the $m_S-r$ plane of possible DM candidates (see text for details). {Here, $m_S$ denotes the mass of the real scalar dark matter candidate $S$, and $r=m_\psi/m_S$ parametrizes the mass splitting between the vector-like fermionic mediator $\psi$ and $S$.}}
\label{fig:results_thermal_plane}
\end{figure}

We show in Fig.~\ref{fig:results_thermal_plane} our limits in the plane spanned by the dark matter mass and the mass ratio, {with the upper (lower) panel corresponding to the scenario of dark matter coupling to $u_R$ ($d_R$)}. The light blue region enclosed by the solid blue line is excluded by our reanalysis of the  ATLAS data, adopting $K=1$ in Eq.~(\ref{eq:sigNLO}); the blue dashed line corresponds to $K=0.5$, and the blue dotted line to $K=2$. Our results {for the case of dark matter coupling to $u_R$} show  that the ATLAS data rules out dark matter masses below $\sim 800$ GeV for $r\gtrsim 1.5$, upon imposing our perturbativity condition $y<6$. This conclusion, furthermore, is rather insensitive to the concrete choice of the $K$ factor. {The ATLAS constraints for the scenario of coupling to $d_R$ are slightly weaker, due to the smaller parton distribution function of the down-quark in the proton, compared to the up-quark.} We also show for completeness the limits from production of vector-like quarks at the Z resonance at LEPI (yellow dotted) and at LEPII (yellow solid). For the former, we imposed $m_\psi \gtrsim m_Z/2$. For the latter, we assumed that the constraints are essentially kinematical and amount to $m_\psi \gtrsim 100$ GeV, similar to LEPII results on squarks searches~\cite{Abbiendi:2002mp}. Besides, we also include in the plots, as a light red region enclosed by a solid red line, the limits from the LUX experiment, discussed in Section \ref{sec:direct-detect-constr}. The complementarity of these two different search strategies in probing the parameter space of the model is manifest from the figure: collider experiments currently provide the best limits for $r\gtrsim 1.4$, while direct detection experiments, for smaller mass splittings. In particular, the null results from the ATLAS experiment rules out the region of ``maximal isospin violation'' at $r\simeq 2.6$ for a DM coupling to $u_R$ ($r\simeq 3.0$ for a DM coupling to $d_R$), for which the sensitivity of xenon-based direct detection experiments is the lowest. We also show, as a dashed red line, the projected reach of the upcoming XENON1T experiment, which will probe dark matter masses in the TeV range for small values of $r$. A future 100 TeV proton-proton collider would perfectly complement the search for large $r$, and in combination with XENON1T, should be able to close in on the parameter space of the model for dark matter masses as heavy as a few tens of TeV.
  
\section{Indirect  detection  constraints}
\label{sec:antipr-constr}

Indirect dark matter searches, using in particular gamma-rays and antiprotons as messengers, provide a complementary avenue to probe our model.  

\begin{figure}[h!]
\hspace*{-0.7cm}  
\begin{tabular}{cc}
    \includegraphics[width=8.8cm]{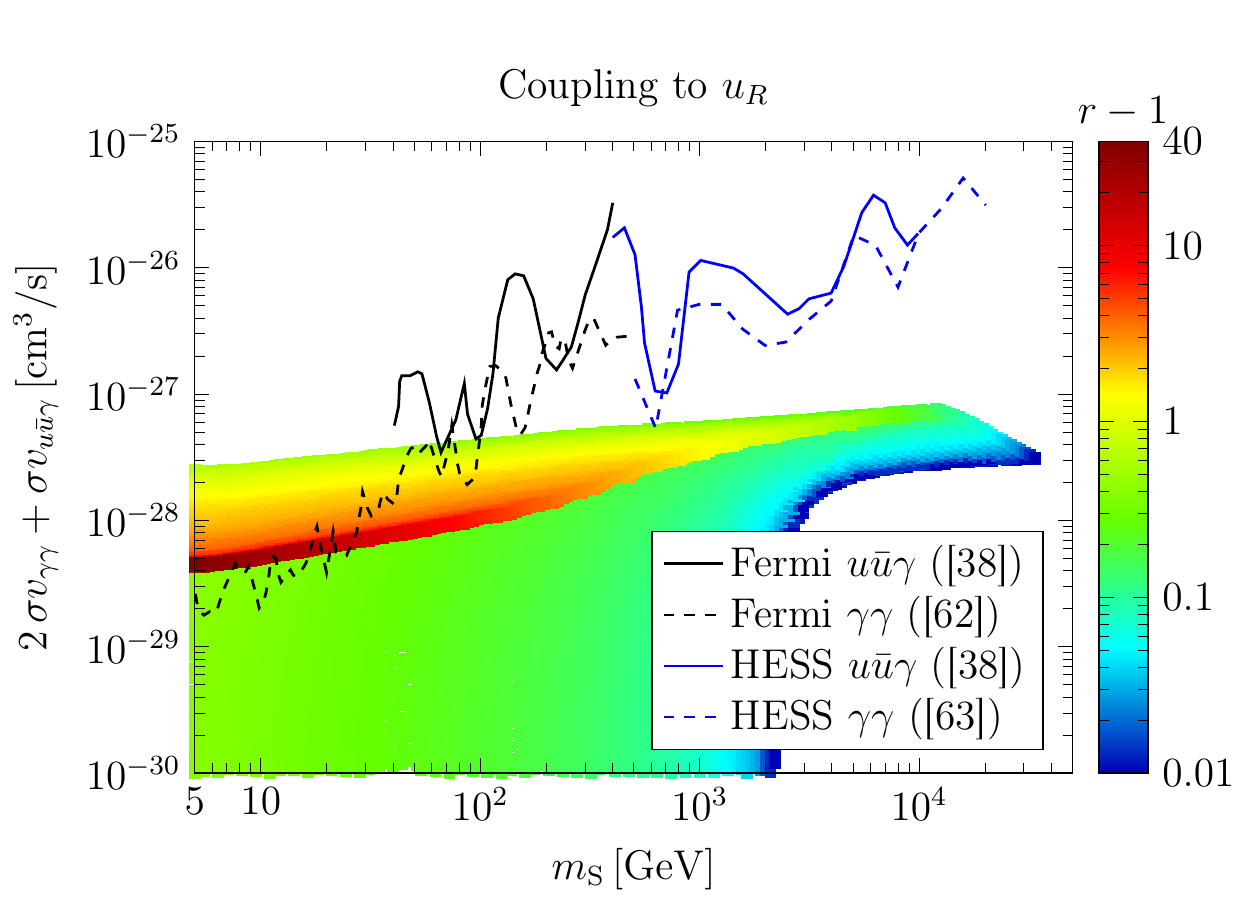} &
    \includegraphics[width=8.8cm]{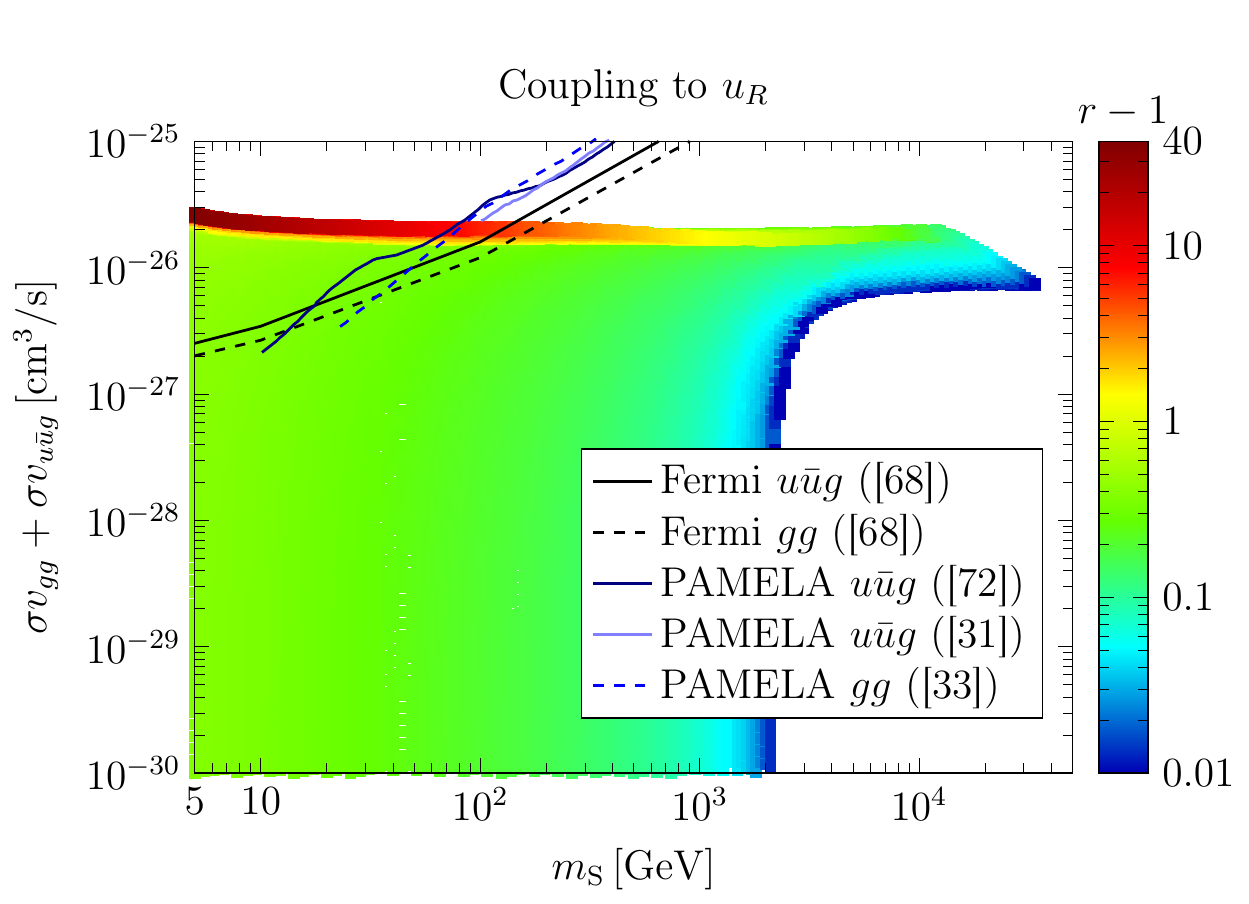}
  \end{tabular}
  \caption{Cross section $\sigma v_{u \bar u\gamma}+ 2 \sigma v_{\gamma\gamma}$ (left panel) and $ \sigma v_{u \bar u g}+ \sigma v_{gg}$ (right panel) expected for thermally produced dark matter as a function of the dark matter mass; the color code indicates different values of $r$. The left panel also shows the upper limits on $\sigma v_{u \bar u\gamma}$ and $ 2 \sigma v_{\gamma\gamma}$ from  the non-observation of sharp spectral features in the  Fermi-LAT and H.E.S.S. data, while the right panel, the limits on $ \sigma v_{u \bar u g}$ and $\sigma v_{gg}$ from the non-observation by the Fermi-LAT of a gamma-ray flux from dSphs, as well as from the non-observation by the PAMELA experiment of an antiproton excess.\label{fig:dsphs}}
\end{figure}

The model under consideration produces gamma-ray signals in the form of sharp spectral features or in the form of a continuum. The sharp spectral features arise mainly from the annihilation final states $u\bar u \gamma$ and $\gamma\gamma$ (the contribution to the total spectrum from $\gamma Z$ is a factor $\tan^2\theta_W\simeq 0.30$ smaller and will be neglected here). The photon multiplicities are analogous to those produced when the dark matter couples to a lepton, which were discussed in \cite{Toma:2013bka,Giacchino:2013bta,Giacchino:2014moa,Ibarra:2014qma}. The final state $u\bar u \gamma$ gives the dominant contribution to the energy spectrum  when the scalar DM and the fermion mediator are very close in mass, while $\gamma\gamma$ {is more important} in the opposite case; for intermediate regimes, the spectrum will be a linear combination of these two, weighted by the corresponding branching fractions. Therefore, for a given dark matter mass, the observational upper limits on $u\bar u \gamma$  and $\gamma\gamma$ bracket the upper limit on $\sigma_{u\bar u \gamma} + 2 \sigma_{\gamma\gamma}$ for any value of the fermion mediator mass. The total annihilation section into sharp gamma-ray spectral features, $\sigma_{u\bar u \gamma} + 2 \sigma_{\gamma\gamma}$, is shown in Fig.~\ref{fig:dsphs}, left panel, in the parameter space spanned by  $m_S$ and $r-1 = m_\psi/m_S-1$, together with the band bracketing the excluded values for this quantity. The band is bounded by the limits on $\sigma_{u \bar u \gamma}$ which were derived in Ref.~\cite{Garny:2013ama} employing  Fermi-LAT data in the energy range 5 - 300 GeV~\cite{Ackermann:2015lka} and H.E.S.S. data in the energy range 500 GeV - 25 TeV~\cite{Abramowski:2013ax}, as well as by the limits on $2 \sigma_{\gamma\gamma}$ derived by the Fermi-LAT collaboration in~\cite{Ackermann:2015lka} in the energy range 200 MeV - 500 GeV and by the H.E.S.S. collaboration in~\cite{Abramowski:2013ax} in the energy range 500 GeV - 25 TeV. In the plot it is assumed an Einasto profile~\cite{Navarro:2003ew,Graham:2005xx,Navarro:2008kc}, with $\alpha=0.17$ and $r_s=20~\rm{kpc}$~\cite{Navarro:2003ew}. {It follows from Fig.~\ref{fig:dsphs} that  the present Fermi-LAT limits on sharp gamma-ray spectral features probe the thermal parameter space of the model for dark matter masses smaller than a few tens of GeV.} {Here, for simplicity we only show the case of dark matter coupling to $u_R$; for coupling to $d_R$ the cross sections predicted by the model are smaller by a factor of a few due to the smaller charge of the $d$-quark compared to the $u$-quark, leading qualitatively to a similar conclusion.}

\begin{figure}[h!] 
\hspace*{-0.7cm}  
\begin{tabular}{cc}
    \includegraphics[width=8.8cm]{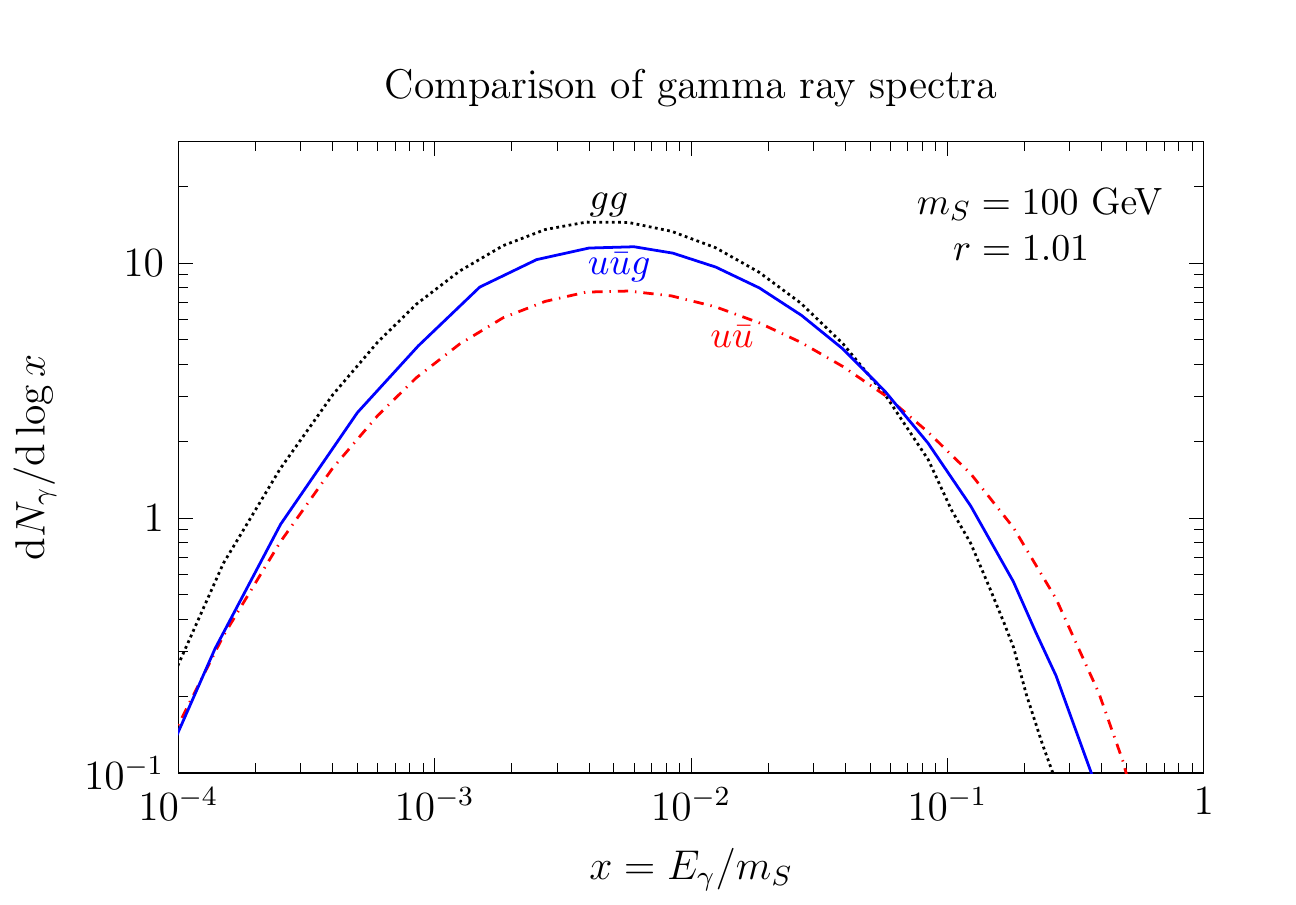} &
    \includegraphics[width=8.8cm]{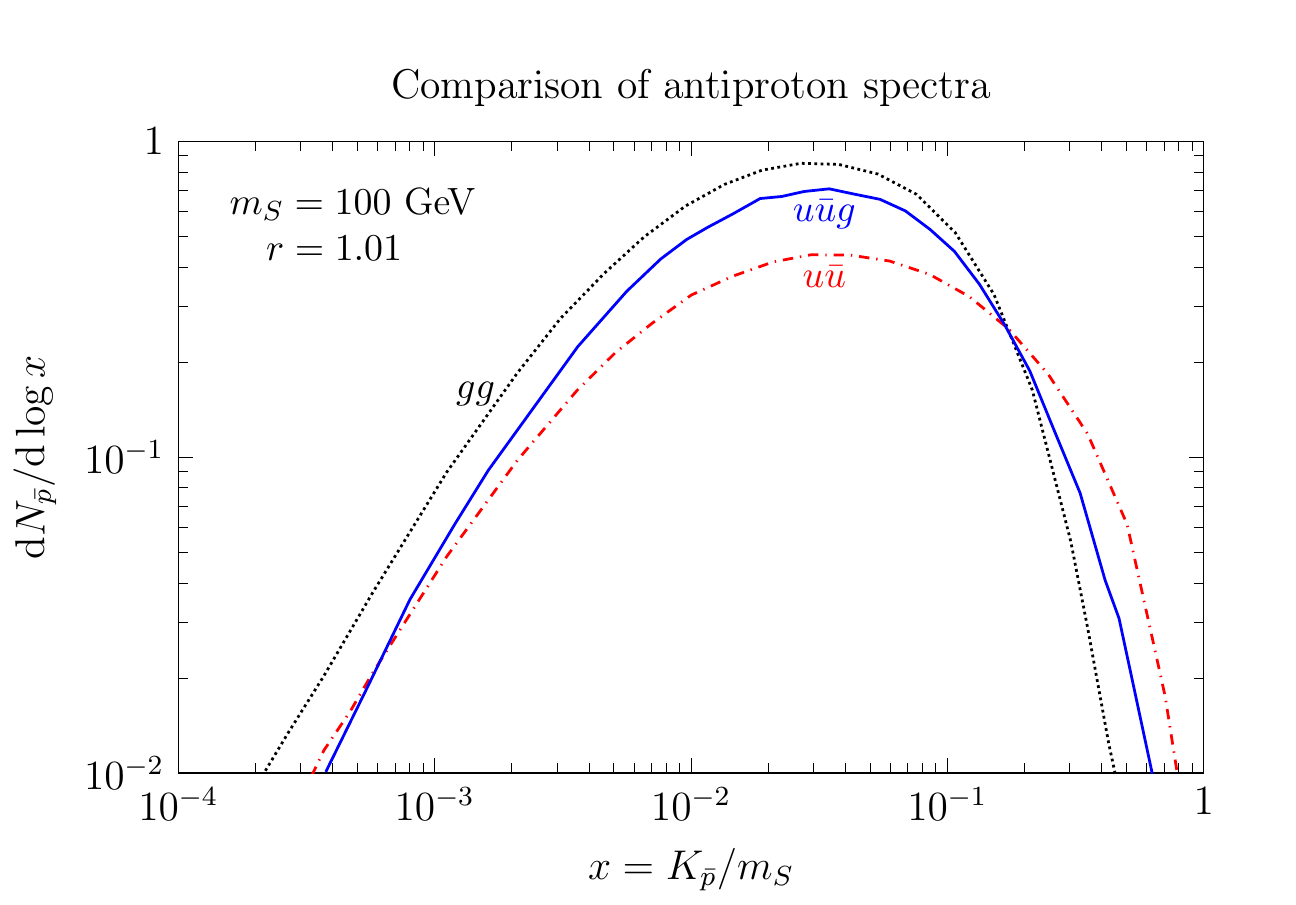}
  \end{tabular}
  \caption{\small  Gamma-ray spectrum (left panel) and antiproton spectrum (right panel) produced by the dark matter annihilation into $gg$ (dotted black line) and $u\bar u g$ (solid blue line), assuming for concreteness $m_S=100$ GeV and $r=1.01$. The figure also shows for comparison the corresponding spectra for the final state $u\bar u$ (dotted-dashed red line).
\label{fig:gammaP}}
\end{figure}

On the other hand, the continuum of gamma-rays is mainly produced by the final states $u\bar u g$ and $gg$ (we neglect in our analysis the contributions from $u\bar u\gamma$ and $\gamma Z$). Similarly to the discussion above, for small mass ratios the energy spectrum of continuum photons is dominated by annihilations into $u\bar u g$, while for large mass ratios, by annihilations into $gg$; the corresponding spectra from annihilation in these two extreme cases are shown in Fig.~\ref{fig:gammaP}.~\footnote{We also show in the same figure the gamma-ray spectrum from the $u\bar u$ final state. The VIB differential cross section is peaked toward $E_g \sim E_{u/\bar u} \sim m_S$ at the parton level. This explains why the spectrum in gamma-rays and anti-protons is roughly half-way between that from $gg$ and $u\bar u$ in Figs.~\ref{fig:gammaP} (see also Ref.~\cite{Bringmann:2015cpa}). This is illustrated for a compressed mass spectrum ($r=1.01$) but we have checked that this feature also holds for larger values of $r$.}  
Again, for intermediate regimes the photon spectrum is a combination of these two, and therefore one can bracket the upper limit on the cross section for any mediator mass by considering the upper limits on the channels $u\bar u g$ and $gg$.  The total annihilation cross section for these two processes, $ \sigma_{u \bar u g} + \sigma_{gg}$, is shown in Fig.~\ref{fig:dsphs}, right panel, together with the band bracketing the upper limits on $\sigma_{\bar u u g} + \sigma_{gg}$ from the Fermi-LAT observations of dSphs~\cite{Ackermann:2015zua}, and which is bounded by the upper limits on $\sigma_{\bar u u g}$ and  $\sigma_{gg}$.~\footnote{The Fermi-LAT collaboration has not published limits on these two final states, therefore, we estimate these limits  following the methodology pursued in~\cite{Bringmann:2012vr}, using the 95 \% CL exclusion limits on DM annihilation into $q \bar q$ from ~\cite{Ackermann:2015zua}. Concretely,  we use the fact that the total gamma-ray flux, {\em e.g.} for the $u \bar u$ channel,  is proportional to $\sigma v_{u \bar u} N_\gamma^{u \bar u}$, where $N_\gamma^{u \bar u}$ is the number of gamma rays per annihilation within the energy range of Fermi-LAT, $0.5$ GeV$ < E_\gamma < 500$ GeV.  We then rescale the  Fermi-LAT limits on  $\sigma v_{u\bar u}$ {to get $\sigma v_{gg,u \bar u g} = \sigma v_{u \bar u} N_\gamma^{u \bar{u}}/N_\gamma^{gg,u \bar u g}$, where we have determined the number of gamma-rays $N_\gamma^{u \bar u,gg,u \bar u g}$ using {\tt PYTHIA 8.1}~\cite{Sjostrand:2007gs}.}} {As can be seen from the Figure, }{the Fermi-LAT limits on continuum gamma-ray emission from dSphs put constraints on the parameter space of the model for $m_S \lesssim 150$ GeV.}

The annihilation channels $u\bar u g$ and $gg$ also produce an antiproton flux, which is currently constrained by the PAMELA data~\cite{Adriani:2012paa} (preliminary data from AMS-02 have been presented in \cite{AMS-days}). We show in Fig.~\ref{fig:dsphs}, right panel, the band bracketing the upper limit on  $\sigma_{\bar u u g} + \sigma_{gg}$ using the limits on $\sigma_{\bar u u g}$  derived in \cite{Asano:2011ik,Garny:2011ii} and the limits on  $\sigma_{\bar g g}$  derived in \cite{Chu:2012qy}. It follows from the Figure that the antiprotons may probe thermal candidates with a mass up to~$\sim 100$~GeV; {it should be borne in mind, though, that the antiproton limits are rather sensitive to uncertainties related to cosmic-ray propagation (these uncertainty may be reduced when more cosmic ray data will be released by the AMS-02 collaboration).}

{Finally, we confront in Fig.~\ref{fig:results_thermal_plane} the limits from indirect detection with those from other search strategies on the thermal dark matter parameter space of the model. For the sake of clarity, and to avoid cluttering of lines in the Figure, we only show the limits from dSphs, which are the strongest and the most robust among the indirect detection constraints analyzed in this paper. As apparent from the Figure, these limits are complementary to those from direct detection and colliders, and confirm, using a completely different search strategy, the exclusion of dark matter particles with mass $m_S\lesssim 150$ GeV and $m_\psi/m_S\gtrsim 1.4$.}


\section{Conclusion}
\label{sec:concl}

In this paper we have studied a simplified dark matter scenario in
which a real scalar dark matter candidate $S$ couples to light
standard model quarks through a Yukawa interaction involving a new
vector-like fermion mediator $\psi$. In order to compute the relic
dark matter abundance via thermal freeze-out, we have taken into
account annihilation into three body final states involving internal
Bremsstrahlung as well as loop processes. We have also accounted for
Sommerfeld corrections of the co-annihilation processes. The latter
give rise up to a 15\% change in the relic density (enhancing or
decreasing the one calculated at perturbative level). In addition, a
distinctive feature of our dark matter scenario is that annihilations
into $gg$ and $q \bar q g$ can be the dominant contributions to the dark
matter annihilations both in the early universe and today.

We have investigated three complementary roads in order to test our
dark matter scenarios: direct detection dark matter searches, collider
searches and indirect detection through annihilation into antiprotons {and gamma rays}.
Current constraints from the LUX experiment exclude most of the viable
parameter space below masses of a hundreds of GeV (as well as a small
island in the TeV range). The reach of the future XENON1T experiment extends to 10 TeV for low values of $m_\psi/m_S-1$.
In addition, we have analyzed how the production of the fermionic
mediator at colliders could complement such constraints. {To this end}, we made use of an ATLAS search for 2-6 jets +
missing $E_T$ using $pp$ collision
data at $\sqrt{s}=8$ TeV with ${\cal L} =20.3$ fb. In order to compute the efficiencies corresponding to our particle
physics model we have simulated events for the production of mediator
pairs using {\tt Madgraph} and applied the cuts corresponding to the
ATLAS search using {\tt CheckMATE}. Our results are the following: for
masses between~$\sim$~30~(100)~GeV and 1 TeV, such collider searches
allow us to probe viable dark matter scenarios that are beyond the
reach of current (future) direct detection searches. Combining the
multijet + missing $E_T$ ATLAS search limit with LUX constraints, real scalar
dark matter coupling to $u$ {or $d$} quarks is excluded for masses in the range
$5-300$ GeV. Moreover, the ATLAS limit
extends up to $m_S\sim 1$ TeV for $m_\psi/m_S-1 \sim {\cal O} (1)$. 
{Finally, we have also estimated the limits from indirect
detection experiments using gamma-rays and antiprotons as messengers.
We have found that the most stringent limits on the model stem from the 
non-observation of a gamma-ray flux correlated to the direction of
dSphs. These limits are complementary to those 
from direct detection and collider experiments and exclude the region 
$m_\psi/m_S\gtrsim 1.4$, $m_S\lesssim 150$ GeV.

\section*{Acknowledgements}
We would like to thank K.~Mawatari, K.~De Causmaecker and M.~Vereecken for helpful discussions on collider constraints and the
use of {\tt MadGraph}, and F.~Maltoni for interesting comments.
 LLH was supported in part by FWO-Vlaanderen
with the post doctoral fellowship project 1271513 and the project
G020714N, by the Belgian Federal Science Policy Office through the
Interuniversity Attraction Pole P7/37 and by the Vrije Universiteit
Brussel through the Strategic Research Program “High-Energy Physics.” The work of F.G. and M.T. is supported by the
Belgian Federal Science Policy Office through the
Interuniversity Attraction Pole P7/37. 
The work of AI and SW has been partially supported by the DFG cluster of excellence ``Origin and Structure of the Universe'', the TUM Graduate School and the Studienstiftung des Deutschen Volkes.

\bibliographystyle{apsrev.bst}
\bibliography{bibcol}

\begin{thebibliography}{72}
\expandafter\ifx\csname natexlab\endcsname\relax\def\natexlab#1{#1}\fi
\expandafter\ifx\csname bibnamefont\endcsname\relax
  \def\bibnamefont#1{#1}\fi
\expandafter\ifx\csname bibfnamefont\endcsname\relax
  \def\bibfnamefont#1{#1}\fi
\expandafter\ifx\csname citenamefont\endcsname\relax
  \def\citenamefont#1{#1}\fi
\expandafter\ifx\csname url\endcsname\relax
  \def\url#1{\texttt{#1}}\fi
\expandafter\ifx\csname urlprefix\endcsname\relax\def\urlprefix{URL }\fi
\providecommand{\bibinfo}[2]{#2}
\providecommand{\eprint}[2][]{\url{#2}}

\bibitem[{\citenamefont{Silveira and Zee}(1985)}]{Silveira:1985rk}
\bibinfo{author}{\bibfnamefont{V.}~\bibnamefont{Silveira}} \bibnamefont{and}
  \bibinfo{author}{\bibfnamefont{A.}~\bibnamefont{Zee}},
  \bibinfo{journal}{Phys.Lett.} \textbf{\bibinfo{volume}{B161}},
  \bibinfo{pages}{136} (\bibinfo{year}{1985}).

\bibitem[{\citenamefont{McDonald}(1994)}]{McDonald:1993ex}
\bibinfo{author}{\bibfnamefont{J.}~\bibnamefont{McDonald}},
  \bibinfo{journal}{Phys.Rev.} \textbf{\bibinfo{volume}{D50}},
  \bibinfo{pages}{3637} (\bibinfo{year}{1994}), \eprint{hep-ph/0702143}.

\bibitem[{\citenamefont{Burgess et~al.}(2001)\citenamefont{Burgess, Pospelov,
  and ter Veldhuis}}]{Burgess:2000yq}
\bibinfo{author}{\bibfnamefont{C.}~\bibnamefont{Burgess}},
  \bibinfo{author}{\bibfnamefont{M.}~\bibnamefont{Pospelov}}, \bibnamefont{and}
  \bibinfo{author}{\bibfnamefont{T.}~\bibnamefont{ter Veldhuis}},
  \bibinfo{journal}{Nucl.Phys.} \textbf{\bibinfo{volume}{B619}},
  \bibinfo{pages}{709} (\bibinfo{year}{2001}), \eprint{hep-ph/0011335}.

\bibitem[{\citenamefont{Patt and Wilczek}(2006)}]{Patt:2006fw}
\bibinfo{author}{\bibfnamefont{B.}~\bibnamefont{Patt}} \bibnamefont{and}
  \bibinfo{author}{\bibfnamefont{F.}~\bibnamefont{Wilczek}}
  (\bibinfo{year}{2006}), \eprint{hep-ph/0605188}.

\bibitem[{\citenamefont{Kadastik et~al.}(2010)\citenamefont{Kadastik, Kannike,
  and Raidal}}]{Kadastik:2009dj}
\bibinfo{author}{\bibfnamefont{M.}~\bibnamefont{Kadastik}},
  \bibinfo{author}{\bibfnamefont{K.}~\bibnamefont{Kannike}}, \bibnamefont{and}
  \bibinfo{author}{\bibfnamefont{M.}~\bibnamefont{Raidal}},
  \bibinfo{journal}{Phys. Rev.} \textbf{\bibinfo{volume}{D81}},
  \bibinfo{pages}{015002} (\bibinfo{year}{2010}), \eprint{0903.2475}.

\bibitem[{\citenamefont{Frigerio and Hambye}(2010)}]{Frigerio:2009wf}
\bibinfo{author}{\bibfnamefont{M.}~\bibnamefont{Frigerio}} \bibnamefont{and}
  \bibinfo{author}{\bibfnamefont{T.}~\bibnamefont{Hambye}},
  \bibinfo{journal}{Phys. Rev.} \textbf{\bibinfo{volume}{D81}},
  \bibinfo{pages}{075002} (\bibinfo{year}{2010}), \eprint{0912.1545}.

\bibitem[{\citenamefont{Arbelaez et~al.}(2015)\citenamefont{Arbelaez, Longas,
  Restrepo, and Zapata}}]{Arbelaez:2015ila}
\bibinfo{author}{\bibfnamefont{C.}~\bibnamefont{Arbelaez}},
  \bibinfo{author}{\bibfnamefont{R.}~\bibnamefont{Longas}},
  \bibinfo{author}{\bibfnamefont{D.}~\bibnamefont{Restrepo}}, \bibnamefont{and}
  \bibinfo{author}{\bibfnamefont{O.}~\bibnamefont{Zapata}}
  (\bibinfo{year}{2015}), \eprint{1509.06313}.

\bibitem[{\citenamefont{Rodejohann and Yaguna}(2015)}]{Rodejohann:2015lca}
\bibinfo{author}{\bibfnamefont{W.}~\bibnamefont{Rodejohann}} \bibnamefont{and}
  \bibinfo{author}{\bibfnamefont{C.~E.} \bibnamefont{Yaguna}}
  (\bibinfo{year}{2015}), \eprint{1509.04036}.

\bibitem[{\citenamefont{Nagata et~al.}(2015)\citenamefont{Nagata, Olive, and
  Zheng}}]{Nagata:2015dma}
\bibinfo{author}{\bibfnamefont{N.}~\bibnamefont{Nagata}},
  \bibinfo{author}{\bibfnamefont{K.~A.} \bibnamefont{Olive}}, \bibnamefont{and}
  \bibinfo{author}{\bibfnamefont{J.}~\bibnamefont{Zheng}}
  (\bibinfo{year}{2015}), \eprint{1509.00809}.

\bibitem[{\citenamefont{Mambrini et~al.}(2015)\citenamefont{Mambrini, Nagata,
  Olive, Quevillon, and Zheng}}]{Mambrini:2015vna}
\bibinfo{author}{\bibfnamefont{Y.}~\bibnamefont{Mambrini}},
  \bibinfo{author}{\bibfnamefont{N.}~\bibnamefont{Nagata}},
  \bibinfo{author}{\bibfnamefont{K.~A.} \bibnamefont{Olive}},
  \bibinfo{author}{\bibfnamefont{J.}~\bibnamefont{Quevillon}},
  \bibnamefont{and} \bibinfo{author}{\bibfnamefont{J.}~\bibnamefont{Zheng}},
  \bibinfo{journal}{Phys. Rev.} \textbf{\bibinfo{volume}{D91}},
  \bibinfo{pages}{095010} (\bibinfo{year}{2015}), \eprint{1502.06929}.

\bibitem[{\citenamefont{Heeck and Patra}(2015)}]{Heeck:2015qra}
\bibinfo{author}{\bibfnamefont{J.}~\bibnamefont{Heeck}} \bibnamefont{and}
  \bibinfo{author}{\bibfnamefont{S.}~\bibnamefont{Patra}},
  \bibinfo{journal}{Phys. Rev. Lett.} \textbf{\bibinfo{volume}{115}},
  \bibinfo{pages}{121804} (\bibinfo{year}{2015}), \eprint{1507.01584}.

\bibitem[{\citenamefont{Goudelis et~al.}(2009)\citenamefont{Goudelis, Mambrini,
  and Yaguna}}]{Goudelis:2009zz}
\bibinfo{author}{\bibfnamefont{A.}~\bibnamefont{Goudelis}},
  \bibinfo{author}{\bibfnamefont{Y.}~\bibnamefont{Mambrini}}, \bibnamefont{and}
  \bibinfo{author}{\bibfnamefont{C.}~\bibnamefont{Yaguna}},
  \bibinfo{journal}{JCAP} \textbf{\bibinfo{volume}{0912}}, \bibinfo{pages}{008}
  (\bibinfo{year}{2009}), \eprint{0909.2799}.

\bibitem[{\citenamefont{Yaguna}(2009)}]{Yaguna:2008hd}
\bibinfo{author}{\bibfnamefont{C.~E.} \bibnamefont{Yaguna}},
  \bibinfo{journal}{JCAP} \textbf{\bibinfo{volume}{0903}}, \bibinfo{pages}{003}
  (\bibinfo{year}{2009}), \eprint{0810.4267}.

\bibitem[{\citenamefont{Gonderinger et~al.}(2010)\citenamefont{Gonderinger, Li,
  Patel, and Ramsey-Musolf}}]{Gonderinger:2009jp}
\bibinfo{author}{\bibfnamefont{M.}~\bibnamefont{Gonderinger}},
  \bibinfo{author}{\bibfnamefont{Y.}~\bibnamefont{Li}},
  \bibinfo{author}{\bibfnamefont{H.}~\bibnamefont{Patel}}, \bibnamefont{and}
  \bibinfo{author}{\bibfnamefont{M.~J.} \bibnamefont{Ramsey-Musolf}},
  \bibinfo{journal}{JHEP} \textbf{\bibinfo{volume}{01}}, \bibinfo{pages}{053}
  (\bibinfo{year}{2010}), \eprint{0910.3167}.

\bibitem[{\citenamefont{Profumo et~al.}(2010)\citenamefont{Profumo, Ubaldi, and
  Wainwright}}]{Profumo:2010kp}
\bibinfo{author}{\bibfnamefont{S.}~\bibnamefont{Profumo}},
  \bibinfo{author}{\bibfnamefont{L.}~\bibnamefont{Ubaldi}}, \bibnamefont{and}
  \bibinfo{author}{\bibfnamefont{C.}~\bibnamefont{Wainwright}},
  \bibinfo{journal}{Phys. Rev.} \textbf{\bibinfo{volume}{D82}},
  \bibinfo{pages}{123514} (\bibinfo{year}{2010}), \eprint{1009.5377}.

\bibitem[{\citenamefont{Djouadi et~al.}(2012)\citenamefont{Djouadi, Lebedev,
  Mambrini, and Quevillon}}]{Djouadi:2011aa}
\bibinfo{author}{\bibfnamefont{A.}~\bibnamefont{Djouadi}},
  \bibinfo{author}{\bibfnamefont{O.}~\bibnamefont{Lebedev}},
  \bibinfo{author}{\bibfnamefont{Y.}~\bibnamefont{Mambrini}}, \bibnamefont{and}
  \bibinfo{author}{\bibfnamefont{J.}~\bibnamefont{Quevillon}},
  \bibinfo{journal}{Phys. Lett.} \textbf{\bibinfo{volume}{B709}},
  \bibinfo{pages}{65} (\bibinfo{year}{2012}), \eprint{1112.3299}.

\bibitem[{\citenamefont{Cline et~al.}(2013)\citenamefont{Cline, Kainulainen,
  Scott, and Weniger}}]{Cline:2013gha}
\bibinfo{author}{\bibfnamefont{J.~M.} \bibnamefont{Cline}},
  \bibinfo{author}{\bibfnamefont{K.}~\bibnamefont{Kainulainen}},
  \bibinfo{author}{\bibfnamefont{P.}~\bibnamefont{Scott}}, \bibnamefont{and}
  \bibinfo{author}{\bibfnamefont{C.}~\bibnamefont{Weniger}},
  \bibinfo{journal}{Phys.Rev.} \textbf{\bibinfo{volume}{D88}},
  \bibinfo{pages}{055025} (\bibinfo{year}{2013}), \eprint{1306.4710}.

\bibitem[{\citenamefont{Duerr et~al.}(2015)\citenamefont{Duerr, Fileviez~Perez,
  and Smirnov}}]{Duerr:2015aka}
\bibinfo{author}{\bibfnamefont{M.}~\bibnamefont{Duerr}},
  \bibinfo{author}{\bibfnamefont{P.}~\bibnamefont{Fileviez~Perez}},
  \bibnamefont{and} \bibinfo{author}{\bibfnamefont{J.}~\bibnamefont{Smirnov}}
  (\bibinfo{year}{2015}), \eprint{1509.04282}.

\bibitem[{\citenamefont{Bertone et~al.}(2009)\citenamefont{Bertone, Jackson,
  Shaughnessy, Tait, and Vallinotto}}]{Bertone:2009cb}
\bibinfo{author}{\bibfnamefont{G.}~\bibnamefont{Bertone}},
  \bibinfo{author}{\bibfnamefont{C.}~\bibnamefont{Jackson}},
  \bibinfo{author}{\bibfnamefont{G.}~\bibnamefont{Shaughnessy}},
  \bibinfo{author}{\bibfnamefont{T.~M.} \bibnamefont{Tait}}, \bibnamefont{and}
  \bibinfo{author}{\bibfnamefont{A.}~\bibnamefont{Vallinotto}},
  \bibinfo{journal}{Phys.Rev.} \textbf{\bibinfo{volume}{D80}},
  \bibinfo{pages}{023512} (\bibinfo{year}{2009}), \eprint{0904.1442}.

\bibitem[{\citenamefont{Boehm and Fayet}(2004)}]{Boehm:2003hm}
\bibinfo{author}{\bibfnamefont{C.}~\bibnamefont{Boehm}} \bibnamefont{and}
  \bibinfo{author}{\bibfnamefont{P.}~\bibnamefont{Fayet}},
  \bibinfo{journal}{Nucl. Phys.} \textbf{\bibinfo{volume}{B683}},
  \bibinfo{pages}{219} (\bibinfo{year}{2004}), \eprint{hep-ph/0305261}.

\bibitem[{\citenamefont{Vasquez et~al.}(2011)\citenamefont{Vasquez, Boehm, and
  Idarraga}}]{Vasquez:2009kq}
\bibinfo{author}{\bibfnamefont{D.~A.} \bibnamefont{Vasquez}},
  \bibinfo{author}{\bibfnamefont{C.}~\bibnamefont{Boehm}}, \bibnamefont{and}
  \bibinfo{author}{\bibfnamefont{J.}~\bibnamefont{Idarraga}},
  \bibinfo{journal}{Phys. Rev.} \textbf{\bibinfo{volume}{D83}},
  \bibinfo{pages}{115017} (\bibinfo{year}{2011}), \eprint{0912.5373}.

\bibitem[{\citenamefont{Fileviez~Perez and Wise}(2013)}]{Perez:2013nra}
\bibinfo{author}{\bibfnamefont{P.}~\bibnamefont{Fileviez~Perez}}
  \bibnamefont{and} \bibinfo{author}{\bibfnamefont{M.~B.} \bibnamefont{Wise}},
  \bibinfo{journal}{JHEP} \textbf{\bibinfo{volume}{1305}}, \bibinfo{pages}{094}
  (\bibinfo{year}{2013}), \eprint{1303.1452}.

\bibitem[{\citenamefont{Toma}(2013)}]{Toma:2013bka}
\bibinfo{author}{\bibfnamefont{T.}~\bibnamefont{Toma}},
  \bibinfo{journal}{Phys.Rev.Lett.} \textbf{\bibinfo{volume}{111}},
  \bibinfo{pages}{091301} (\bibinfo{year}{2013}), \eprint{1307.6181}.

\bibitem[{\citenamefont{Giacchino et~al.}(2013)\citenamefont{Giacchino,
  Lopez-Honorez, and Tytgat}}]{Giacchino:2013bta}
\bibinfo{author}{\bibfnamefont{F.}~\bibnamefont{Giacchino}},
  \bibinfo{author}{\bibfnamefont{L.}~\bibnamefont{Lopez-Honorez}},
  \bibnamefont{and} \bibinfo{author}{\bibfnamefont{M.~H.}
  \bibnamefont{Tytgat}}, \bibinfo{journal}{JCAP}
  \textbf{\bibinfo{volume}{1310}}, \bibinfo{pages}{025} (\bibinfo{year}{2013}),
  \eprint{1307.6480}.

\bibitem[{\citenamefont{Chang et~al.}(2014{\natexlab{a}})\citenamefont{Chang,
  Edezhath, Hutchinson, and Luty}}]{Chang:2014tea}
\bibinfo{author}{\bibfnamefont{S.}~\bibnamefont{Chang}},
  \bibinfo{author}{\bibfnamefont{R.}~\bibnamefont{Edezhath}},
  \bibinfo{author}{\bibfnamefont{J.}~\bibnamefont{Hutchinson}},
  \bibnamefont{and} \bibinfo{author}{\bibfnamefont{M.}~\bibnamefont{Luty}},
  \bibinfo{journal}{Phys. Rev.} \textbf{\bibinfo{volume}{D90}},
  \bibinfo{pages}{015011} (\bibinfo{year}{2014}{\natexlab{a}}),
  \eprint{1402.7358}.

\bibitem[{\citenamefont{Ibarra et~al.}(2014{\natexlab{a}})\citenamefont{Ibarra,
  Toma, Totzauer, and Wild}}]{Ibarra:2014qma}
\bibinfo{author}{\bibfnamefont{A.}~\bibnamefont{Ibarra}},
  \bibinfo{author}{\bibfnamefont{T.}~\bibnamefont{Toma}},
  \bibinfo{author}{\bibfnamefont{M.}~\bibnamefont{Totzauer}}, \bibnamefont{and}
  \bibinfo{author}{\bibfnamefont{S.}~\bibnamefont{Wild}},
  \bibinfo{journal}{Phys. Rev.} \textbf{\bibinfo{volume}{D90}},
  \bibinfo{pages}{043526} (\bibinfo{year}{2014}{\natexlab{a}}),
  \eprint{1405.6917}.

\bibitem[{\citenamefont{Giacchino et~al.}(2014)\citenamefont{Giacchino,
  Lopez-Honorez, and Tytgat}}]{Giacchino:2014moa}
\bibinfo{author}{\bibfnamefont{F.}~\bibnamefont{Giacchino}},
  \bibinfo{author}{\bibfnamefont{L.}~\bibnamefont{Lopez-Honorez}},
  \bibnamefont{and} \bibinfo{author}{\bibfnamefont{M.~H.~G.}
  \bibnamefont{Tytgat}}, \bibinfo{journal}{JCAP}
  \textbf{\bibinfo{volume}{1408}}, \bibinfo{pages}{046} (\bibinfo{year}{2014}),
  \eprint{1405.6921}.

\bibitem[{\citenamefont{Bergstrom}(1989)}]{Bergstrom:1989jr}
\bibinfo{author}{\bibfnamefont{L.}~\bibnamefont{Bergstrom}},
  \bibinfo{journal}{Phys.Lett.} \textbf{\bibinfo{volume}{B225}},
  \bibinfo{pages}{372} (\bibinfo{year}{1989}).

\bibitem[{\citenamefont{Flores et~al.}(1989)\citenamefont{Flores, Olive, and
  Rudaz}}]{Flores:1989ru}
\bibinfo{author}{\bibfnamefont{R.}~\bibnamefont{Flores}},
  \bibinfo{author}{\bibfnamefont{K.~A.} \bibnamefont{Olive}}, \bibnamefont{and}
  \bibinfo{author}{\bibfnamefont{S.}~\bibnamefont{Rudaz}},
  \bibinfo{journal}{Phys.Lett.} \textbf{\bibinfo{volume}{B232}},
  \bibinfo{pages}{377} (\bibinfo{year}{1989}).

\bibitem[{\citenamefont{Bringmann et~al.}(2008)\citenamefont{Bringmann,
  Bergstrom, and Edsjo}}]{Bringmann:2007nk}
\bibinfo{author}{\bibfnamefont{T.}~\bibnamefont{Bringmann}},
  \bibinfo{author}{\bibfnamefont{L.}~\bibnamefont{Bergstrom}},
  \bibnamefont{and} \bibinfo{author}{\bibfnamefont{J.}~\bibnamefont{Edsjo}},
  \bibinfo{journal}{JHEP} \textbf{\bibinfo{volume}{0801}}, \bibinfo{pages}{049}
  (\bibinfo{year}{2008}), \eprint{0710.3169}.

\bibitem[{\citenamefont{Garny et~al.}(2012)\citenamefont{Garny, Ibarra, and
  Vogl}}]{Garny:2011ii}
\bibinfo{author}{\bibfnamefont{M.}~\bibnamefont{Garny}},
  \bibinfo{author}{\bibfnamefont{A.}~\bibnamefont{Ibarra}}, \bibnamefont{and}
  \bibinfo{author}{\bibfnamefont{S.}~\bibnamefont{Vogl}},
  \bibinfo{journal}{JCAP} \textbf{\bibinfo{volume}{1204}}, \bibinfo{pages}{033}
  (\bibinfo{year}{2012}), \eprint{1112.5155}.

\bibitem[{\citenamefont{Bringmann et~al.}(2012)\citenamefont{Bringmann, Huang,
  Ibarra, Vogl, and Weniger}}]{Bringmann:2012vr}
\bibinfo{author}{\bibfnamefont{T.}~\bibnamefont{Bringmann}},
  \bibinfo{author}{\bibfnamefont{X.}~\bibnamefont{Huang}},
  \bibinfo{author}{\bibfnamefont{A.}~\bibnamefont{Ibarra}},
  \bibinfo{author}{\bibfnamefont{S.}~\bibnamefont{Vogl}}, \bibnamefont{and}
  \bibinfo{author}{\bibfnamefont{C.}~\bibnamefont{Weniger}},
  \bibinfo{journal}{JCAP} \textbf{\bibinfo{volume}{1207}}, \bibinfo{pages}{054}
  (\bibinfo{year}{2012}), \eprint{1203.1312}.

\bibitem[{\citenamefont{Chu et~al.}(2012)\citenamefont{Chu, Hambye, Scarna, and
  Tytgat}}]{Chu:2012qy}
\bibinfo{author}{\bibfnamefont{X.}~\bibnamefont{Chu}},
  \bibinfo{author}{\bibfnamefont{T.}~\bibnamefont{Hambye}},
  \bibinfo{author}{\bibfnamefont{T.}~\bibnamefont{Scarna}}, \bibnamefont{and}
  \bibinfo{author}{\bibfnamefont{M.~H.~G.} \bibnamefont{Tytgat}},
  \bibinfo{journal}{Phys. Rev.} \textbf{\bibinfo{volume}{D86}},
  \bibinfo{pages}{083521} (\bibinfo{year}{2012}), \eprint{1206.2279}.

\bibitem[{\citenamefont{Garny et~al.}(2014)\citenamefont{Garny, Ibarra,
  Rydbeck, and Vogl}}]{Garny:2014waa}
\bibinfo{author}{\bibfnamefont{M.}~\bibnamefont{Garny}},
  \bibinfo{author}{\bibfnamefont{A.}~\bibnamefont{Ibarra}},
  \bibinfo{author}{\bibfnamefont{S.}~\bibnamefont{Rydbeck}}, \bibnamefont{and}
  \bibinfo{author}{\bibfnamefont{S.}~\bibnamefont{Vogl}},
  \bibinfo{journal}{JHEP} \textbf{\bibinfo{volume}{06}}, \bibinfo{pages}{169}
  (\bibinfo{year}{2014}), \eprint{1403.4634}.

\bibitem[{\citenamefont{Akerib et~al.}(2014)}]{Akerib:2013tjd}
\bibinfo{author}{\bibfnamefont{D.~S.} \bibnamefont{Akerib}}
  \bibnamefont{et~al.} (\bibinfo{collaboration}{LUX}), \bibinfo{journal}{Phys.
  Rev. Lett.} \textbf{\bibinfo{volume}{112}}, \bibinfo{pages}{091303}
  (\bibinfo{year}{2014}), \eprint{1310.8214}.

\bibitem[{\citenamefont{Aprile}(2013)}]{Aprile:2012zx}
\bibinfo{author}{\bibfnamefont{E.}~\bibnamefont{Aprile}}
  (\bibinfo{collaboration}{XENON1T}), \bibinfo{journal}{Springer Proc. Phys.}
  \textbf{\bibinfo{volume}{148}}, \bibinfo{pages}{93} (\bibinfo{year}{2013}),
  \eprint{1206.6288}.

\bibitem[{\citenamefont{Hisano et~al.}(2015)\citenamefont{Hisano, Nagai, and
  Nagata}}]{Hisano:2015bma}
\bibinfo{author}{\bibfnamefont{J.}~\bibnamefont{Hisano}},
  \bibinfo{author}{\bibfnamefont{R.}~\bibnamefont{Nagai}}, \bibnamefont{and}
  \bibinfo{author}{\bibfnamefont{N.}~\bibnamefont{Nagata}},
  \bibinfo{journal}{JHEP} \textbf{\bibinfo{volume}{05}}, \bibinfo{pages}{037}
  (\bibinfo{year}{2015}), \eprint{1502.02244}.

\bibitem[{\citenamefont{Garny et~al.}(2013)\citenamefont{Garny, Ibarra, Pato,
  and Vogl}}]{Garny:2013ama}
\bibinfo{author}{\bibfnamefont{M.}~\bibnamefont{Garny}},
  \bibinfo{author}{\bibfnamefont{A.}~\bibnamefont{Ibarra}},
  \bibinfo{author}{\bibfnamefont{M.}~\bibnamefont{Pato}}, \bibnamefont{and}
  \bibinfo{author}{\bibfnamefont{S.}~\bibnamefont{Vogl}},
  \bibinfo{journal}{JCAP} \textbf{\bibinfo{volume}{1312}}, \bibinfo{pages}{046}
  (\bibinfo{year}{2013}), \eprint{1306.6342}.

\bibitem[{\citenamefont{Ibarra et~al.}(2014{\natexlab{b}})\citenamefont{Ibarra,
  Totzauer, and Wild}}]{Ibarra:2014vya}
\bibinfo{author}{\bibfnamefont{A.}~\bibnamefont{Ibarra}},
  \bibinfo{author}{\bibfnamefont{M.}~\bibnamefont{Totzauer}}, \bibnamefont{and}
  \bibinfo{author}{\bibfnamefont{S.}~\bibnamefont{Wild}},
  \bibinfo{journal}{JCAP} \textbf{\bibinfo{volume}{1404}}, \bibinfo{pages}{012}
  (\bibinfo{year}{2014}{\natexlab{b}}), \eprint{1402.4375}.

\bibitem[{\citenamefont{Kopp et~al.}(2014)\citenamefont{Kopp, Michaels, and
  Smirnov}}]{Kopp:2014tsa}
\bibinfo{author}{\bibfnamefont{J.}~\bibnamefont{Kopp}},
  \bibinfo{author}{\bibfnamefont{L.}~\bibnamefont{Michaels}}, \bibnamefont{and}
  \bibinfo{author}{\bibfnamefont{J.}~\bibnamefont{Smirnov}},
  \bibinfo{journal}{JCAP} \textbf{\bibinfo{volume}{1404}}, \bibinfo{pages}{022}
  (\bibinfo{year}{2014}), \eprint{1401.6457}.

\bibitem[{\citenamefont{Papucci et~al.}(2014)\citenamefont{Papucci, Vichi, and
  Zurek}}]{Papucci:2014iwa}
\bibinfo{author}{\bibfnamefont{M.}~\bibnamefont{Papucci}},
  \bibinfo{author}{\bibfnamefont{A.}~\bibnamefont{Vichi}}, \bibnamefont{and}
  \bibinfo{author}{\bibfnamefont{K.~M.} \bibnamefont{Zurek}},
  \bibinfo{journal}{JHEP} \textbf{\bibinfo{volume}{11}}, \bibinfo{pages}{024}
  (\bibinfo{year}{2014}), \eprint{1402.2285}.

\bibitem[{\citenamefont{Ibarra et~al.}(2015)\citenamefont{Ibarra, Pierce, Shah,
  and Vogl}}]{Ibarra:2015nca}
\bibinfo{author}{\bibfnamefont{A.}~\bibnamefont{Ibarra}},
  \bibinfo{author}{\bibfnamefont{A.}~\bibnamefont{Pierce}},
  \bibinfo{author}{\bibfnamefont{N.~R.} \bibnamefont{Shah}}, \bibnamefont{and}
  \bibinfo{author}{\bibfnamefont{S.}~\bibnamefont{Vogl}},
  \bibinfo{journal}{Phys. Rev.} \textbf{\bibinfo{volume}{D91}},
  \bibinfo{pages}{095018} (\bibinfo{year}{2015}), \eprint{1501.03164}.

\bibitem[{\citenamefont{Garny et~al.}(2015)\citenamefont{Garny, Ibarra, and
  Vogl}}]{Garny:2015wea}
\bibinfo{author}{\bibfnamefont{M.}~\bibnamefont{Garny}},
  \bibinfo{author}{\bibfnamefont{A.}~\bibnamefont{Ibarra}}, \bibnamefont{and}
  \bibinfo{author}{\bibfnamefont{S.}~\bibnamefont{Vogl}}
  (\bibinfo{year}{2015}), \eprint{1503.01500}.

\bibitem[{\citenamefont{Bélanger et~al.}(2015)\citenamefont{Bélanger,
  Boudjema, Pukhov, and Semenov}}]{Belanger:2014vza}
\bibinfo{author}{\bibfnamefont{G.}~\bibnamefont{Bélanger}},
  \bibinfo{author}{\bibfnamefont{F.}~\bibnamefont{Boudjema}},
  \bibinfo{author}{\bibfnamefont{A.}~\bibnamefont{Pukhov}}, \bibnamefont{and}
  \bibinfo{author}{\bibfnamefont{A.}~\bibnamefont{Semenov}},
  \bibinfo{journal}{Comput. Phys. Commun.} \textbf{\bibinfo{volume}{192}},
  \bibinfo{pages}{322} (\bibinfo{year}{2015}), \eprint{1407.6129}.

\bibitem[{\citenamefont{Griest and Seckel}(1991)}]{Griest:1990kh}
\bibinfo{author}{\bibfnamefont{K.}~\bibnamefont{Griest}} \bibnamefont{and}
  \bibinfo{author}{\bibfnamefont{D.}~\bibnamefont{Seckel}},
  \bibinfo{journal}{Phys.Rev.} \textbf{\bibinfo{volume}{D43}},
  \bibinfo{pages}{3191} (\bibinfo{year}{1991}).

\bibitem[{\citenamefont{De~Simone et~al.}(2014)\citenamefont{De~Simone,
  Giudice, and Strumia}}]{deSimone:2014pda}
\bibinfo{author}{\bibfnamefont{A.}~\bibnamefont{De~Simone}},
  \bibinfo{author}{\bibfnamefont{G.~F.} \bibnamefont{Giudice}},
  \bibnamefont{and} \bibinfo{author}{\bibfnamefont{A.}~\bibnamefont{Strumia}},
  \bibinfo{journal}{JHEP} \textbf{\bibinfo{volume}{06}}, \bibinfo{pages}{081}
  (\bibinfo{year}{2014}), \eprint{1402.6287}.

\bibitem[{\citenamefont{Cassel}(2010)}]{Cassel:2009wt}
\bibinfo{author}{\bibfnamefont{S.}~\bibnamefont{Cassel}}, \bibinfo{journal}{J.
  Phys.} \textbf{\bibinfo{volume}{G37}}, \bibinfo{pages}{105009}
  (\bibinfo{year}{2010}), \eprint{0903.5307}.

\bibitem[{\citenamefont{Iengo}(2009)}]{Iengo:2009ni}
\bibinfo{author}{\bibfnamefont{R.}~\bibnamefont{Iengo}},
  \bibinfo{journal}{JHEP} \textbf{\bibinfo{volume}{05}}, \bibinfo{pages}{024}
  (\bibinfo{year}{2009}), \eprint{0902.0688}.

\bibitem[{\citenamefont{Drees and Nojiri}(1993)}]{Drees:1993bu}
\bibinfo{author}{\bibfnamefont{M.}~\bibnamefont{Drees}} \bibnamefont{and}
  \bibinfo{author}{\bibfnamefont{M.}~\bibnamefont{Nojiri}},
  \bibinfo{journal}{Phys. Rev.} \textbf{\bibinfo{volume}{D48}},
  \bibinfo{pages}{3483} (\bibinfo{year}{1993}), \eprint{hep-ph/9307208}.

\bibitem[{\citenamefont{Hisano et~al.}(2010)\citenamefont{Hisano, Ishiwata, and
  Nagata}}]{Hisano:2010ct}
\bibinfo{author}{\bibfnamefont{J.}~\bibnamefont{Hisano}},
  \bibinfo{author}{\bibfnamefont{K.}~\bibnamefont{Ishiwata}}, \bibnamefont{and}
  \bibinfo{author}{\bibfnamefont{N.}~\bibnamefont{Nagata}},
  \bibinfo{journal}{Phys. Rev.} \textbf{\bibinfo{volume}{D82}},
  \bibinfo{pages}{115007} (\bibinfo{year}{2010}), \eprint{1007.2601}.

\bibitem[{\citenamefont{Akerib et~al.}(2015)}]{Akerib:2015rjg}
\bibinfo{author}{\bibfnamefont{D.~S.} \bibnamefont{Akerib}}
  \bibnamefont{et~al.} (\bibinfo{collaboration}{LUX}) (\bibinfo{year}{2015}),
  \eprint{1512.03506}.

\bibitem[{\citenamefont{Billard et~al.}(2014)\citenamefont{Billard, Strigari,
  and Figueroa-Feliciano}}]{Billard:2013qya}
\bibinfo{author}{\bibfnamefont{J.}~\bibnamefont{Billard}},
  \bibinfo{author}{\bibfnamefont{L.}~\bibnamefont{Strigari}}, \bibnamefont{and}
  \bibinfo{author}{\bibfnamefont{E.}~\bibnamefont{Figueroa-Feliciano}},
  \bibinfo{journal}{Phys. Rev.} \textbf{\bibinfo{volume}{D89}},
  \bibinfo{pages}{023524} (\bibinfo{year}{2014}), \eprint{1307.5458}.

\bibitem[{\citenamefont{Belyaev et~al.}(2013)\citenamefont{Belyaev,
  Christensen, and Pukhov}}]{Belyaev:2012qa}
\bibinfo{author}{\bibfnamefont{A.}~\bibnamefont{Belyaev}},
  \bibinfo{author}{\bibfnamefont{N.~D.} \bibnamefont{Christensen}},
  \bibnamefont{and} \bibinfo{author}{\bibfnamefont{A.}~\bibnamefont{Pukhov}},
  \bibinfo{journal}{Comput. Phys. Commun.} \textbf{\bibinfo{volume}{184}},
  \bibinfo{pages}{1729} (\bibinfo{year}{2013}), \eprint{1207.6082}.

\bibitem[{\citenamefont{Kidonakis}(2010)}]{Kidonakis:2010dk}
\bibinfo{author}{\bibfnamefont{N.}~\bibnamefont{Kidonakis}},
  \bibinfo{journal}{Phys. Rev.} \textbf{\bibinfo{volume}{D82}},
  \bibinfo{pages}{114030} (\bibinfo{year}{2010}), \eprint{1009.4935}.

\bibitem[{\citenamefont{Chang et~al.}(2014{\natexlab{b}})\citenamefont{Chang,
  Edezhath, Hutchinson, and Luty}}]{Chang:2013oia}
\bibinfo{author}{\bibfnamefont{S.}~\bibnamefont{Chang}},
  \bibinfo{author}{\bibfnamefont{R.}~\bibnamefont{Edezhath}},
  \bibinfo{author}{\bibfnamefont{J.}~\bibnamefont{Hutchinson}},
  \bibnamefont{and} \bibinfo{author}{\bibfnamefont{M.}~\bibnamefont{Luty}},
  \bibinfo{journal}{Phys.Rev.} \textbf{\bibinfo{volume}{D89}},
  \bibinfo{pages}{015011} (\bibinfo{year}{2014}{\natexlab{b}}),
  \eprint{1307.8120}.

\bibitem[{\citenamefont{collaboration}(2013)}]{TheATLAScollaboration:2013fha}
\bibinfo{author}{\bibfnamefont{T.~A.} \bibnamefont{collaboration}}
  (\bibinfo{collaboration}{ATLAS}) (\bibinfo{year}{2013}).

\bibitem[{\citenamefont{Alloul et~al.}(2014)\citenamefont{Alloul, Christensen,
  Degrande, Duhr, and Fuks}}]{Alloul:2013bka}
\bibinfo{author}{\bibfnamefont{A.}~\bibnamefont{Alloul}},
  \bibinfo{author}{\bibfnamefont{N.~D.} \bibnamefont{Christensen}},
  \bibinfo{author}{\bibfnamefont{C.}~\bibnamefont{Degrande}},
  \bibinfo{author}{\bibfnamefont{C.}~\bibnamefont{Duhr}}, \bibnamefont{and}
  \bibinfo{author}{\bibfnamefont{B.}~\bibnamefont{Fuks}},
  \bibinfo{journal}{Comput. Phys. Commun.} \textbf{\bibinfo{volume}{185}},
  \bibinfo{pages}{2250} (\bibinfo{year}{2014}), \eprint{1310.1921}.

\bibitem[{\citenamefont{Alwall et~al.}(2014)\citenamefont{Alwall, Frederix,
  Frixione, Hirschi, Maltoni, Mattelaer, Shao, Stelzer, Torrielli, and
  Zaro}}]{Alwall:2014hca}
\bibinfo{author}{\bibfnamefont{J.}~\bibnamefont{Alwall}},
  \bibinfo{author}{\bibfnamefont{R.}~\bibnamefont{Frederix}},
  \bibinfo{author}{\bibfnamefont{S.}~\bibnamefont{Frixione}},
  \bibinfo{author}{\bibfnamefont{V.}~\bibnamefont{Hirschi}},
  \bibinfo{author}{\bibfnamefont{F.}~\bibnamefont{Maltoni}},
  \bibinfo{author}{\bibfnamefont{O.}~\bibnamefont{Mattelaer}},
  \bibinfo{author}{\bibfnamefont{H.~S.} \bibnamefont{Shao}},
  \bibinfo{author}{\bibfnamefont{T.}~\bibnamefont{Stelzer}},
  \bibinfo{author}{\bibfnamefont{P.}~\bibnamefont{Torrielli}},
  \bibnamefont{and} \bibinfo{author}{\bibfnamefont{M.}~\bibnamefont{Zaro}},
  \bibinfo{journal}{JHEP} \textbf{\bibinfo{volume}{07}}, \bibinfo{pages}{079}
  (\bibinfo{year}{2014}), \eprint{1405.0301}.

\bibitem[{\citenamefont{Sjostrand et~al.}(2006)\citenamefont{Sjostrand, Mrenna,
  and Skands}}]{Sjostrand:2006za}
\bibinfo{author}{\bibfnamefont{T.}~\bibnamefont{Sjostrand}},
  \bibinfo{author}{\bibfnamefont{S.}~\bibnamefont{Mrenna}}, \bibnamefont{and}
  \bibinfo{author}{\bibfnamefont{P.~Z.} \bibnamefont{Skands}},
  \bibinfo{journal}{JHEP} \textbf{\bibinfo{volume}{05}}, \bibinfo{pages}{026}
  (\bibinfo{year}{2006}), \eprint{hep-ph/0603175}.

\bibitem[{\citenamefont{Drees et~al.}(2014)\citenamefont{Drees, Dreiner,
  Schmeier, Tattersall, and Kim}}]{Drees:2013wra}
\bibinfo{author}{\bibfnamefont{M.}~\bibnamefont{Drees}},
  \bibinfo{author}{\bibfnamefont{H.}~\bibnamefont{Dreiner}},
  \bibinfo{author}{\bibfnamefont{D.}~\bibnamefont{Schmeier}},
  \bibinfo{author}{\bibfnamefont{J.}~\bibnamefont{Tattersall}},
  \bibnamefont{and} \bibinfo{author}{\bibfnamefont{J.~S.} \bibnamefont{Kim}},
  \bibinfo{journal}{Comput. Phys. Commun.} \textbf{\bibinfo{volume}{187}},
  \bibinfo{pages}{227} (\bibinfo{year}{2014}), \eprint{1312.2591}.

\bibitem[{\citenamefont{Abbiendi et~al.}(2002)}]{Abbiendi:2002mp}
\bibinfo{author}{\bibfnamefont{G.}~\bibnamefont{Abbiendi}} \bibnamefont{et~al.}
  (\bibinfo{collaboration}{OPAL}), \bibinfo{journal}{Phys. Lett.}
  \textbf{\bibinfo{volume}{B545}}, \bibinfo{pages}{272} (\bibinfo{year}{2002}),
  \bibinfo{note}{[Erratum: Phys. Lett.B548,258(2002)]},
  \eprint{hep-ex/0209026}.

\bibitem[{\citenamefont{Ackermann
  et~al.}(2015{\natexlab{a}})}]{Ackermann:2015lka}
\bibinfo{author}{\bibfnamefont{M.}~\bibnamefont{Ackermann}}
  \bibnamefont{et~al.} (\bibinfo{collaboration}{Fermi-LAT}),
  \bibinfo{journal}{Phys. Rev.} \textbf{\bibinfo{volume}{D91}},
  \bibinfo{pages}{122002} (\bibinfo{year}{2015}{\natexlab{a}}),
  \eprint{1506.00013}.

\bibitem[{\citenamefont{Abramowski et~al.}(2013)}]{Abramowski:2013ax}
\bibinfo{author}{\bibfnamefont{A.}~\bibnamefont{Abramowski}}
  \bibnamefont{et~al.} (\bibinfo{collaboration}{H.E.S.S. Collaboration}),
  \bibinfo{journal}{Phys.Rev.Lett.} \textbf{\bibinfo{volume}{110}},
  \bibinfo{pages}{041301} (\bibinfo{year}{2013}), \eprint{1301.1173}.

\bibitem[{\citenamefont{Navarro et~al.}(2004)\citenamefont{Navarro, Hayashi,
  Power, Jenkins, Frenk et~al.}}]{Navarro:2003ew}
\bibinfo{author}{\bibfnamefont{J.~F.} \bibnamefont{Navarro}},
  \bibinfo{author}{\bibfnamefont{E.}~\bibnamefont{Hayashi}},
  \bibinfo{author}{\bibfnamefont{C.}~\bibnamefont{Power}},
  \bibinfo{author}{\bibfnamefont{A.}~\bibnamefont{Jenkins}},
  \bibinfo{author}{\bibfnamefont{C.~S.} \bibnamefont{Frenk}},
  \bibnamefont{et~al.}, \bibinfo{journal}{Mon. Not. Roy. Astron. Soc.}
  \textbf{\bibinfo{volume}{349}}, \bibinfo{pages}{1039} (\bibinfo{year}{2004}),
  \eprint{astro-ph/0311231}.

\bibitem[{\citenamefont{Graham et~al.}(2006)\citenamefont{Graham, Merritt,
  Moore, Diemand, and Terzic}}]{Graham:2005xx}
\bibinfo{author}{\bibfnamefont{A.~W.} \bibnamefont{Graham}},
  \bibinfo{author}{\bibfnamefont{D.}~\bibnamefont{Merritt}},
  \bibinfo{author}{\bibfnamefont{B.}~\bibnamefont{Moore}},
  \bibinfo{author}{\bibfnamefont{J.}~\bibnamefont{Diemand}}, \bibnamefont{and}
  \bibinfo{author}{\bibfnamefont{B.}~\bibnamefont{Terzic}},
  \bibinfo{journal}{Astron.J.} \textbf{\bibinfo{volume}{132}},
  \bibinfo{pages}{2685} (\bibinfo{year}{2006}), \eprint{astro-ph/0509417}.

\bibitem[{\citenamefont{Navarro et~al.}(2008)\citenamefont{Navarro, Ludlow,
  Springel, Wang, Vogelsberger et~al.}}]{Navarro:2008kc}
\bibinfo{author}{\bibfnamefont{J.~F.} \bibnamefont{Navarro}},
  \bibinfo{author}{\bibfnamefont{A.}~\bibnamefont{Ludlow}},
  \bibinfo{author}{\bibfnamefont{V.}~\bibnamefont{Springel}},
  \bibinfo{author}{\bibfnamefont{J.}~\bibnamefont{Wang}},
  \bibinfo{author}{\bibfnamefont{M.}~\bibnamefont{Vogelsberger}},
  \bibnamefont{et~al.} (\bibinfo{year}{2008}), \eprint{0810.1522}.

\bibitem[{\citenamefont{Bringmann et~al.}(2015)\citenamefont{Bringmann, Galea,
  and Walia}}]{Bringmann:2015cpa}
\bibinfo{author}{\bibfnamefont{T.}~\bibnamefont{Bringmann}},
  \bibinfo{author}{\bibfnamefont{A.~J.} \bibnamefont{Galea}}, \bibnamefont{and}
  \bibinfo{author}{\bibfnamefont{P.}~\bibnamefont{Walia}}
  (\bibinfo{year}{2015}), \eprint{1510.02473}.

\bibitem[{\citenamefont{Ackermann
  et~al.}(2015{\natexlab{b}})}]{Ackermann:2015zua}
\bibinfo{author}{\bibfnamefont{M.}~\bibnamefont{Ackermann}}
  \bibnamefont{et~al.} (\bibinfo{collaboration}{Fermi-LAT})
  (\bibinfo{year}{2015}{\natexlab{b}}), \eprint{1503.02641}.

\bibitem[{\citenamefont{Sjostrand et~al.}(2008)\citenamefont{Sjostrand, Mrenna,
  and Skands}}]{Sjostrand:2007gs}
\bibinfo{author}{\bibfnamefont{T.}~\bibnamefont{Sjostrand}},
  \bibinfo{author}{\bibfnamefont{S.}~\bibnamefont{Mrenna}}, \bibnamefont{and}
  \bibinfo{author}{\bibfnamefont{P.~Z.} \bibnamefont{Skands}},
  \bibinfo{journal}{Comput. Phys. Commun.} \textbf{\bibinfo{volume}{178}},
  \bibinfo{pages}{852} (\bibinfo{year}{2008}), \eprint{0710.3820}.

\bibitem[{\citenamefont{Adriani et~al.}(2013)}]{Adriani:2012paa}
\bibinfo{author}{\bibfnamefont{O.}~\bibnamefont{Adriani}} \bibnamefont{et~al.},
  \bibinfo{journal}{JETP Lett.} \textbf{\bibinfo{volume}{96}},
  \bibinfo{pages}{621} (\bibinfo{year}{2013}), \bibinfo{note}{[Pisma Zh. Eksp.
  Teor. Fiz.96,693(2012)]}.

\bibitem[{AMS()}]{AMS-days}
\bibinfo{note}{A. Kounine, \textit{Latest AMS Results: The Positron Fraction
  and the p-bar/p Ratio}. Talk presented at the AMS Days at CERN, Geneva, 15-17
  April 2015.}

\bibitem[{\citenamefont{Asano et~al.}(2012)\citenamefont{Asano, Bringmann, and
  Weniger}}]{Asano:2011ik}
\bibinfo{author}{\bibfnamefont{M.}~\bibnamefont{Asano}},
  \bibinfo{author}{\bibfnamefont{T.}~\bibnamefont{Bringmann}},
  \bibnamefont{and} \bibinfo{author}{\bibfnamefont{C.}~\bibnamefont{Weniger}},
  \bibinfo{journal}{Phys. Lett.} \textbf{\bibinfo{volume}{B709}},
  \bibinfo{pages}{128} (\bibinfo{year}{2012}), \eprint{1112.5158}.

\end{thebibliography}

\end{document}